\documentclass[%
 reprint,
superscriptaddress,
nofootinbib,
 amsmath,amssymb,
 aps,floatfix,
]{revtex4-1}

\usepackage{amsmath,amssymb}
\usepackage{graphicx}
\usepackage{dcolumn}
\usepackage{bm}
\usepackage{hyperref}
\usepackage{epsfig}
\input{epsf.sty}
\usepackage{multirow}
\begin{document}

\title{Spinning gravimagnetic particles in Schwarzschild-like black holes}

\author{Walberto Guzm\'an-Ram\'irez}%
\email{ wguzman@fat.uerj.br}
\affiliation{Departamento de Matem\'atica, F\'isica e Computa\c c\~ao - FAT, Universidade do Estado do Rio de Janeiro, RJ, Brazil}

\author{Ricardo Becerril}%
 \email{ricardo.becerril@umich.mx}
 \affiliation{Instituto de F\'isica y Matem\'aticas, Universidad Michoacana de San Nicol\'as de Hidalgo. Edif. C-3, 58040 Morelia, Michoac\'an, M\'exico}
 
 \author{Susana Valdez-Alvarado}%
 \email{svaldeza@uaemex.mx}
\affiliation{Facultad de Ciencias de la Universidad Aut\'onoma del Estado de M\'exico  (UAEM\'ex.), Instituto Literario No. 100, C.P. 50000, Toluca, Estado de M\'exico, M\'exico}

\author{Alexei A. Deriglazov}%
\email{alexei.deriglazov@ufjf.edu.br}
\affiliation{Departamento de Matem\'atica, ICE, Universidade Federal de Juiz de
Fora, MG, Brazil. \\  Department of Physics, Tomsk State University, Lenin Prospect 36, 634050, Tomsk, Russia
}

\date{November 2021}

\begin{abstract}
We study the motion of a spinning particle with gravimagnetic moment in Schwarzschild-like spacetimes with a metric $ds^2=-f(r) dt^2 + f^{-1}(r) dr^2 + r^2 d\Omega^2$, specifically we deal with Schwarzschild, Reissner-Nordstrom black holes as well as Ayon-Beato-Garcia and Bardeen regular spacetimes. First, we introduce the Hamiltonian system of equations which describes such kind of particles. In the case of null gravimagnetic moment, the equations are equivalent to the Mathisson-Papapetrou-Tulczyjew-Dixon (MPTD) equations. Working in the equatorial plane, using the constants of motion generated by the symmetries of the considered spacetimes and the Spin Supplementary Conditions (SSC), we change the problem of solving six differential equations for the momenta and the non-vanishing spin-tensor components to solving six algebraic equations. We show that the equation for the $P_0(r)$ component totally decouples, $P_0(r)$ can be found by solving a 6th order polynomial. We analyze the conditions for existence of solutions of this algebraic system for the relevant  cases of gravimagnetic moment equal to unit, which corresponds to a gravimagnetic particle, and zero which corresponds to the  MTPD system. A numerical algorithm to generate solutions of the momenta $P_{\mu}$ is provided and some solutions are generated. 
\end{abstract}

\maketitle 

\section{Introduction}

The detection of gravitational waves by LIGO and VIRGO detectors  comes after years of technological developments \cite{Abbott1, Abbott2}, confirming one of the most important predictions of General Relativity. Future observations, coming from the existing ground-based detectors and the planned space-based missions such as LISA,  promise to open new windows for physics where the gravitational field is strong. Among the candidates for gravitational waves sources are the extreme mass-ratio inspirals (EMRIs) systems where one component  has negligible mass as compared to the companion.  
 As a first approximation, EMRIs are modeled by a point test particle in motion around an isolated black hole (BH). Despite the simplification, this approximation gives essential information about the motion of the bodies orbiting a supermassive object.  
A most realistic model, in the test particle approximation,  should consider internal degrees of freedom of the test body, such that spin or internal rotation. The interaction between the spin and the gravitational field can significantly modify the trajectories and precessions of the test bodies, reflecting in the forms of gravitational waves.
Besides the observational importance,  the study of relativistic  spinning particles in gravitational fields has importance from the theoretical point of view. For instance,  Wald \cite{Wald} showed  that the interaction between the spin and the gravitational field of a spherical body with a slow rotation is similar to the electromagnetic  interaction between two dipoles.  This result was recently extended to the case when the particle has gravimagnetic moment \cite{Deriglazov:2018vwa, Ramirez:2017pmp}.  
The conventional equations used to describe a spinning object moving in a gravitational field,  are the manifestly generally covariant Mathisson-Papapetrou-Dixon equations 
\cite{Mathisson:1937zz, Fock1939, Papapetrou:1951pa, Tulc, Dixon1964}.  These equations are formulated assuming that the test body structure could be described by a set of multipoles and that the approximation involves only the first two terms (the pole-dipole approximation). The equations are then derived by the integration of the conservation law for the energy-momentum tensor: $T^{\mu\nu}_{\ \ ;\mu}=0$. Those equations are
\begin{eqnarray}
\nabla p_\mu &&= -\frac 12 R_{\mu\nu\alpha\beta} u^\nu s^{\alpha\beta} \, , \label{MPD1}\\
\nabla s^{\mu\nu}&& = p^\mu u^\nu - p^\nu u^\mu \, \label{MPD2} \, , 
\end{eqnarray}
where $\tau$, $u^\mu = \frac{dx^\mu}{d\tau}$,  $p^\mu$, $s^{\mu\nu}$ are the proper time,  the four-velocity of the particle,  its momentum and  the spin tensor, respectively. In addition to (\ref{MPD1}) and (\ref{MPD2}), it is necessary to include some conditions on the spin's components, the so-called SSC. In this paper, we are interested in the Tulczyjew SSC
\begin{equation}
p_\mu s^{\mu\nu} =0 \, . 
\label{T-c}
\end{equation}
This condition reduces the number of independent components of $s^{\mu\nu}$. Additionally, it permits to establish a relation between $u^\mu$ and $p^\mu$.  
We refer the equations (\ref{MPD1})-(\ref{MPD2}) together with (\ref{T-c}) as Mathisson-Papapetrou-Tulczyjew-Dixon (MPTD) equations.
The r.h.s. of (\ref{MPD1}) is the so called spin-orbit coupling. Here we shall study the  higher coupling (\ref{g-m}), called gravimagnetic moment.
This paper is organized as follows. The dynamical equations describing a spinning particle with gravimagnetic moment are established in section \ref{HESP}. In section \ref{S-spacetime}, we study the motion of spinning particles in a spacetime given by
$ds^2=-f(r) c^2 dt^2+ \frac{dr^2}{f(r)} +r^2(d\theta^2+\sin^2{\theta} d\phi^2)$
particularly, we consider the Schwarzschild and Reissner-Nordstrom black holes as well as the regular Ayon-Beato-Garcia and Bardeen regular black holes. In section IV, we provide explicit expression to relate velocities and momenta for both cases: MPTD and gravimagnetic particles. Conclusions and some final remarks are provided in section V. \\

{\bf Notation.} Our variables employ an arbitrary parametrization $\sigma$, then 
$\dot x^\mu=\frac{dx^\mu}{d\sigma}$. Square brackets stand for antisymmetrization: 
$P^{[\mu}\dot x^{\nu]}=P^\mu \dot x^{\nu} - P^\nu \dot x^\mu$. 
Notation for the scalar functions constructed from four-dimensional vectors and second-rank tensors are $Z^2=g_{\mu\nu}Z^\mu Z^\nu$,  $\dot x^\mu G_{\mu\nu}\dot x^\nu=\dot x\cdot G \cdot \dot x$,   $\theta \cdot S=\theta^{\mu\nu}S_{\mu\nu}$,  $\mu, \nu=0,1, 2, 3$.
%
\section{Hamiltonian system of equations for spinning particles}\label{HESP}

The dynamics of a spinning particle in curved spacetimes  can be obtained from a variational formalism \cite{Hojman:1978wk, Rietdijk:1992tx, DPW2, DW2015.1, Ramirez:2017pmp, Porto, Barausse}. In this paper, we shall use the equations derived from  the vector models of spin \cite{DPM1, DPW2, Deriglazov:2017jub, DWGR2016, DW2015.1, Ramirez:2017pmp}. In these models, the basic variables used to introduce the spin are the non-Grassmann vector $\omega^\mu$ and its conjugate momentum $\pi_\mu$, being the spin tensor a composed quantity constructed from these variables $S^{\mu\nu}= 2(\omega^\mu \pi^\nu - \omega^\nu \pi^\mu)$. Using this formalism in \cite{Ramirez:2017pmp} we introduce the non-minimal interaction
\begin{equation}
 \frac{\kappa}{16}\theta \cdot S \, , \label{g-m}
 \end{equation}
where the tensor $\theta_{\mu\nu}\equiv R_{\mu\nu\alpha\beta} S^{\alpha\beta}$ is the gravitational analogy of the electromagnetic field strength.
That coupling  is similar to the interaction of a spinning particle with an electromagnetic field through the gyromagnetic ratio $g$.
Given this similarity, (\ref{g-m}) is called  gravimagnetic interaction \cite{Pomeransky1998, khriplovich:1996, dAmbrosi:2015ndl, Ambrosi,DWGR2016, Deriglazov:2017jub,  Deriglazov:2018vwa, Ramirez:2017pmp}, and the  constant $\kappa$ is called gravimagnetic moment. A spinning particle with interaction (\ref{g-m}) has a correct ultrarelativistic acceleration behavior when $\kappa=1$, contrary to the one described by the MPTD equations, see \cite{Deriglazov:2017jub, DWGR2016, Deriglazov:2015zta, Deriglazov:2014zzm}.

The Hamiltonian equations, coming from the vector models of spin, for the particle's position $x^\mu(\sigma)$,  canonical momentum $P_\mu$, and the spin tensor are \cite{Ramirez:2017pmp}
\begin{eqnarray}
\nabla P_\mu &=& -\frac 14 \theta_{\mu\nu}\dot x^\nu - \frac{\lambda\kappa}{32} (\nabla_\mu  R_{\alpha\beta\sigma\lambda})
S^{\alpha\beta}S^{\sigma\lambda} \, , \label{motion-P-k} \\
\nabla S^{\mu\nu} &=& 2P^{[\mu} \dot x^{\nu]} +\frac{\lambda\kappa}{4}\theta^{[\mu}{}_{\alpha} S^{\nu]\alpha}  \,  \label{motion-S-k} \, , \\
\dot x^\mu &=& \lambda  \left[\delta^\mu{}_\nu -\bar a(\kappa -1) S^{\mu\alpha}\theta_{\alpha\nu}\right] P^\nu  
+ \lambda \frac{c \kappa \bar a}{b}  Z^\mu  \, ,
\label{motion-x-k}  
\end{eqnarray}
where
$\nabla$ denotes the covariant derivative along the curve $x^\mu(\sigma)$ ($\nabla P_\mu= \frac{dP_\mu}{d\sigma} -\Gamma^\alpha_{\mu\beta}\dot x^\beta P_\alpha$, $\ldots$).  The vector $Z^\mu$ is defined as
 \begin{equation}
 Z^\mu \equiv \frac{b}{8c} S^{\mu\nu}(\nabla_\nu  R_{\alpha\beta\sigma\lambda}) S^{\alpha\beta}S^{\sigma\lambda} \, , \label {Z-k}
 \end{equation}
and the scalars $\bar a$ and $b$ are 
\begin{eqnarray} 
&& \bar a \equiv \frac{ 2 }{ 16\mu^2+(\kappa+1)(\theta \cdot S) } \, , \label{a-bar} \\
&& b \equiv \frac{1}{8\mu^2+\kappa(\theta\cdot S)} \, ,\label{b}
 \end{eqnarray}
with $\mu^2 = m_0^2c^2$ being $m_0$ the mass of the test particle.
In vector models of spin, besides the equations of motion, we obtain the conditions 
\begin{eqnarray}
&&S^{\mu\nu}P_\nu = 0 \, ,\label{condition1} \\
&&S^{\mu\nu} S_{\mu\nu} = 8\alpha \, ,\label{condition2} \\
&&P^2 + \frac{\kappa}{16} (\theta\cdot S) + \mu^2 = 0 \, . \label{mass-shell}
\end{eqnarray}
These conditions come from the constraints of the model. The first condition is just the Tulczyjew SSC (\ref{T-c}), and the second establishes that the magnitude of the spin is a constant, being  $\alpha$  the value of the spin. The conditions (\ref{condition1}) and (\ref{condition2})  imply that only two components of spin-tensor are independent, as it should be for an elementary spin one-half particle. The last equation is the mass-shell condition of the model. Let us remark that in this model (\ref{condition1})-(\ref{mass-shell}) are not introduced by hand; they  are a consequence of the Dirac  procedure for  singular systems \cite{Dirac, Gitman, deriglazov2010classical}. 

The presence of $\lambda$ in equations  (\ref{motion-P-k})-(\ref{motion-x-k}) is related to the reparameterization invariance of the model. In the following subsections, we will exclude it in the particular cases of gravimagnetic moments equal to one and zero.
\subsection{Gravimagnetic particles $\kappa=1$}
\label{G-particle}
We will refer to the particles with $\kappa=1$  as  gravimagnetic particles. In this case, equation (\ref{motion-x-k}) and condition (\ref{mass-shell}) are  
\begin{eqnarray}
\dot x^\mu= \lambda  \left[P^\mu + c Z^\mu \right] \, ,\label{motion-x-1}  \\
P^2 + \frac{1}{16} (\theta\cdot S) + \mu^2 = 0 \, . \label{ms-1}
\end{eqnarray}
The Lagrangian multiplier $\lambda$, can be excluded using the standard procedure, i.e.,  taking the square of (\ref{motion-x-1})
\begin{eqnarray}
\dot x^\mu g_{\mu\nu} \dot x^\nu &=& \lambda^2\left[P^\mu + c Z^\mu \right] g_{\mu\nu} \left[P^\nu + c Z^\nu \right] \nonumber \\
&=& \lambda^2\left[P^2  +  c^2 Z^2 \right] \, , \label{sl-2}
\end{eqnarray}
where we have used that $P_\mu Z^\mu=0$. Now,  replacing $P^2$ from (\ref{ms-1})  we get
\begin{equation}
\lambda^2 = \frac{ - \dot x^\mu g_{\mu\nu} \dot x^\nu }{m^2_r c^2},  \label{sl3}
\end{equation}
where we have defined the scalar $m_r$ as
\begin{equation}
m_r^2c^2\equiv \mu^2 + \frac{1}{16 } (\theta\cdot S) - c^2 Z^2 \, .
\end{equation}
Substituting (\ref{sl3}) in  (\ref{motion-x-1}) we get
\begin{eqnarray}
\frac{ \dot x^\mu }{ \sqrt{ -\dot x \cdot g \cdot \dot x} } 
 = \frac{1}{ m_r c } \left[ P^\mu + cZ^\mu \right] \, .  \label{motion-x1}
\end{eqnarray} 
The resulting equation of motion is manifestly  invariant under reparameterizations.  In (\ref{motion-x1}), we can choose any parameter to describe the curve, for instance, let us choose  proper time 
\begin{equation}
c^2 d\tau^2 = -g_{\mu\nu} dx^\mu  dx^\nu \label{sl-1} \, .
\end{equation}
For this choice, the equations  (\ref{motion-P-k}), (\ref{motion-S-k}) and (\ref{motion-x1}) become
\begin{eqnarray}
\nabla P_\mu &=& -\frac 14 \theta_{\mu\nu}\dot x^\nu - \frac{1}{32 m_r} (\nabla_\mu  R_{\alpha\beta\sigma\lambda})
S^{\alpha\beta}S^{\sigma\lambda} \, , \label{motion-P-1} \\
\nabla S^{\mu\nu} &=& 2P^{[\mu} \dot x^{\nu]} + \frac{1}{4m_r}\theta^{[\mu}{}_{\alpha} S^{\nu]\alpha}  \,  \label{motion-S-1} \, , \\
\dot x^\mu &=& \frac{1}{m_r}  \left[P^\mu + c Z^\mu \right] \, . \label{sl4}  
\end{eqnarray}
Equations (\ref{motion-P-1})-(\ref{sl4}), together with the SSC (\ref{condition1}) and (\ref{condition2}) and the condition (\ref{ms-1}), form the hamiltonian system of equations describing a gravimagnetic particle.
The scalar $m_r$ in (\ref{sl4}) is a ``constant'' of normalization, i.e., the square of the r.h.s of (\ref{sl4}) is equal to $-c^2$, which is consistent with (\ref{sl-1}). Let us remark that both the vector $Z^\mu$ and the scalar $m_r$ depend on the spin variables and the Riemann tensor and not on the momenta. Then, from (\ref{sl4}), we trivially obtain the momentum-velocity relation
\begin{equation}
P^\mu = m_r \dot x^\mu - c Z^\mu \, .
\end{equation}
This relation can be replaced in (\ref{motion-P-1}) and (\ref{motion-S-1}) resulting the Lagrangian equations of the system, i.e.,   second-order differential  equations for $x^\mu$. 

\subsection{ \texorpdfstring{$\kappa=0$}{k} particles} 

In the case of  $\kappa=0$ equations (\ref{motion-x-k}) and  (\ref{mass-shell}) are
%
\begin{eqnarray}
&&\dot x^\mu= \lambda  T^{\mu}{}_{\nu}  P^\nu   \, , \label{motion-x-0}  \\
&&P^2  + \mu^2 = 0 \, , \label{ms-0}
\end{eqnarray}
where $T^{\mu}{}_{\nu}$ is the second rank tensor defined as
\begin{equation}
T^{\mu}{}_{\nu} = \delta^\mu{}_\nu + \bar a S^{\mu\sigma}\theta_{\sigma\nu} \, , 
\label{sl0-3} 
\end{equation}
with $\bar a =  \frac{ 2 }{ 16\mu^2+ \theta \cdot S } $. 
 Following the same procedure to obtain (\ref{sl3}), but using (\ref{motion-x-0}), at this time we get
 \begin{equation}
 \lambda^2= \frac{ - \dot x^\mu g_{\mu\nu} \dot x^\nu}{  - P \cdot G \cdot P   } \, , \label{sl0-6}
 \end{equation}
 where we have defined the second rank tensor 
 \begin{eqnarray}
  G_{\alpha\beta} \equiv T^{\mu}{}_{\alpha} g_{\mu\nu} T^{\nu}{}_{\beta} \, . \label{sl0-G}
\end{eqnarray}
Substituting (\ref{sl0-6}) in (\ref{motion-x-0}) we get
\begin{equation}
\frac{ \dot x^\mu }{ \sqrt{ -  \dot x \cdot g \cdot \dot x } } = \frac{1}{ \sqrt{ - P \cdot G \cdot P  } } T^{\mu}{}_{\nu} P^\nu \, . \label{sl0-9}
\end{equation}
Once again, the resulting equation of motion is manifestly  invariant under reparameterizations. Choosing the proper time parameterization (\ref{sl-1}), the equations (\ref{motion-P-k}), (\ref{motion-S-k}) and (\ref{motion-x-0})  become 
 \begin{eqnarray}
\nabla P_\mu &=& -\frac 14 \theta_{\mu\nu}\dot x^\nu  \, , \label{motion-P-0} \\
\nabla S^{\mu\nu} &=& 2P^{[\mu} \dot x^{\nu]}   \label{motion-S-0} \, , \\
\dot x^\mu &=& \frac{c}{ \sqrt{ - P \cdot G \cdot P } } T^{\mu}{}_{\nu} P^\nu \, . \label{sl0-7} 
\end{eqnarray}
Equation (\ref{sl0-7}) is normalized to $c^2$, with a  factor of normalization depending on the momenta.
Using the SSC (\ref{condition1}) and (\ref{ms-0}), the scalar into the square root takes the form 
\begin{equation}
- P^\alpha G_{\alpha\beta} P^\beta =  \mu^2 -\frac{ 4 P^\alpha \theta_{\alpha\sigma} S^{\sigma\mu} g_{\mu\nu} S^{\nu\lambda } 
\theta_{\lambda\beta} P^\beta} {(16\mu^2 + \theta\cdot S)^2} .
\end{equation}
Replacing this in (\ref{sl0-7}), we obtain the equation for $u^\mu$ introduced in \cite{Suzuki:1996gm}.  
To obtain the  momentum-velocity relation in this case, we first  multiplied    (\ref{motion-x-0}) by the inverse of $T^{\mu}{}_{\nu}$\footnote{To show that (\ref{T-1}) is the inverse of (\ref{sl0-3}) we use the identity $S^{\mu\alpha} \theta_{\alpha\beta} S^{\beta\nu}=-\frac{1}{2} (\theta \cdot S)S^{\mu\nu}$, which comes from the definition of $S^{\mu\nu}$ in the vector-model of spin} 
\begin{equation}
(T^{-1})^{\mu}{}_{\nu} = \delta^\mu_\nu - \frac{1}{8\mu^2} S^{\mu\alpha}\theta_{\alpha\nu} \, , \label{T-1}
\end{equation}
obtaining 
\begin{equation}
P^\mu = \frac{1}{\lambda} (T^{-1})^{\mu}{}_{\nu} \dot x^\nu \, . \label{P-vel-0}
\end{equation}
Now, using (\ref{ms-0}) we get 
%
\begin{equation}
\lambda^{2} = \frac{ - \dot x \cdot \bar G \cdot \dot x   }{\mu^2} \, , \label{L-0}
\end{equation}
where
\begin{equation}
\quad \bar G_{\mu\nu} \equiv (T^{-1})^{\alpha}{}_{\mu} g_{\alpha\beta} (T^{-1})^{\beta}{}_{\nu} \, .
\end{equation}
Finally, replacing (\ref{L-0}) in (\ref{P-vel-0}) we obtain
\begin{equation}
P^\mu =  \frac{\mu }{ \sqrt{- \dot x \cdot \bar G \cdot \dot x  } } (T^{-1})^{\mu}{}_{\nu} \dot x^\nu \, . \label{P-x-0}
\end{equation}

The r.h.s of this equation does not depend on the momenta, so,  substituting (\ref{P-x-0}) in (\ref{motion-P-0}) and (\ref{motion-S-0}) we obtain the Lagrangian equations for the system. \\
%

{\bf Comparison with MPTD equations}\label{comparation MPTD} \\

We can compare the dynamics of particles with $\kappa=0$, equations (\ref{motion-P-0})-(\ref{sl0-7}) and conditions (\ref{condition1}), (\ref{condition2}) and (\ref{ms-0}) with the dynamics described by the MPTD equations.  The equations (\ref{motion-P-0}) and (\ref{motion-S-0}) together with (\ref{condition1}) form the MPTD equations  (\ref{MPD1})-(\ref{T-c}) (our spin is twice that of Dixon). 
Concerning to conditions (\ref{condition2}) and (\ref{ms-0}), in \cite{DWGR2016} and \cite{DW2015.1}, it is showed that, the MPTD equations  imply that $\sqrt{-p^2}=k=$const. and $s^{\mu\nu}s_{\mu\nu}=\beta=$const., so, the class of trajectories of our spinning particle with $\kappa=0$, $\mu=k$ and $8\alpha=\beta$ corresponds to the MPTD theory.  To close this correspondence,  in the following paragraphs, we will show how (\ref{sl0-7}) is obtained from the MPTD equations.

First, we considered the consistency  of (\ref{T-c}), i.e., $\frac{d}{d\sigma} (p_{\mu} s^{\mu\nu})=0$, from this, we obtain the equation
\begin{equation}
p^2 u^\mu - (p\cdot u) p^\mu + \frac 12 s^{\mu\alpha}\bar\theta_{\alpha\beta} u^\beta =0 \, , \label{MPD4}  
\end{equation}
where $\bar\theta_{\mu\nu}\equiv R_{\mu\nu\alpha\beta} s^{\alpha\beta}$.
Now, let us consider the auxiliary vector $A^\mu$, so that
\begin{equation}
u^\mu = p^\mu + A^\mu \, . \label{MP4}
\end{equation}
Replacing this in  the consistency relation (\ref{MPD4}) we obtain the equation
\begin{equation}
N^\nu{}_\alpha A^\alpha = \frac{ A \cdot p }{p^2} \left( \delta^\nu_\alpha - \frac{1}{2 A \cdot p} s^{\nu\lambda}\bar\theta_{\lambda \alpha} \right) p^\alpha \, , \label{MP6}
\end{equation}
where  the second rank tensor $N^\nu{}_\alpha$ is defined as
\begin{equation}
N^\nu{}_\alpha = \delta^\nu_\alpha + \frac{1}{2p^2} s^{\nu\lambda}\bar\theta_{\lambda \alpha}\, .\label{MP7}
\end{equation}
This tensor possesses  inverse given by
\begin{equation}
\left( N^{-1} \right)^\mu{}_\alpha = \frac{1}{\xi} \left[ \left(  1-\frac{1}{4p^2} \bar\theta\cdot s \right)  \delta^\mu_\alpha - \frac{1}{2p^2} s^{\nu\lambda}\bar\theta_{\lambda \alpha} \right] \, , \label{MP8}
\end{equation}
where\footnote{The symbol ${}^\star$ is used to denote the dual of a skew-symmetric  tensor $A^{\mu\nu}$, defined as
$A^{\star \mu\nu } =\frac 12 \epsilon^{\mu\nu\alpha\beta} A_{\alpha\beta}$. To prove  that (\ref{MP8}) is the inverse of (\ref{MP7}), we use the identities; $A^{\mu\alpha} B_{\alpha\nu} - B^{\star \mu\alpha} A^{\star}{}_{\alpha\nu} \equiv -\frac 12 \delta^{\mu}_\nu (A^{\alpha\beta}B_{\alpha\beta})$ where $A^{\mu\nu}$ and $B^{\mu\nu}$ are two arbitrary skew-symmetric tensors, and the identity $A^{\mu\sigma}A^{\star}{}_{\sigma\nu}= -\frac 14 \delta^\mu_\nu (A^{\alpha\beta}A^{\star}_{\alpha\beta})$ \cite{Obukhov:2010kn}. } 
\begin{equation}
\xi= 1 - \frac{1}{4p^2} (\bar\theta\cdot s) - \frac{1}{4^3p^4} (s\cdot s^{\star})(\bar\theta\cdot \bar\theta^\star) \, . \label{MP9}
\end{equation}
Multiplying  (\ref{MP6}) by $N^{-1}$ we obtain $A^\alpha$, and substituting the result in (\ref{MP4}) we get
\begin{eqnarray}
u^\mu =  \left[ \frac{ p^2 + A\cdot p }{\xi p^4} \right] \ M^\mu{}_\alpha p^\alpha \, ,  \label{MP11}
\end{eqnarray}
where
\begin{eqnarray}
M^\mu{}_\alpha \equiv  \delta^\mu_{\alpha} - \frac{2}{4p^2-\bar\theta\cdot s}s^{\mu\lambda}\bar\theta_{\lambda\alpha}  \, .  \label{M} 
\end{eqnarray}
To exclude $A^\mu$, we take the square of  (\ref{MP11}), obtaining   
 \begin{equation}
  \left[ \frac{ p^2 + A \cdot p }{\xi p^4} \right]^2 = \frac{c^2}{  p \cdot \bar G \cdot p  } \, , \label{MP12}
 \end{equation}
 where 
 \begin{equation}
 \bar G_{\alpha\beta} \equiv M^{\mu}{}_{\alpha} g_{\mu\nu} M^{\nu}{}_{\beta} \, . \label{MP14}
 \end{equation}
 Then, the velocity-momentum relation, consistent with (\ref{MPD1})-(\ref{T-c}) is 
 \begin{equation}
 u^\mu =  \frac{c}{ \sqrt{ - p \cdot \bar G  \cdot p  } }  \ M^\mu{}_\alpha p^\alpha \, .\label{MP15}
 \end{equation}
As we mentioned $p^2$ is a constant in the MPTD theory,  so, considering the mass-shell condition $p^2 +\mu^2=0$ (i.e. $k=\mu$),  $M^\mu{}_\alpha$ becomes
\begin{equation}
M^\mu{}_\alpha = \delta^\mu_{\alpha} + \frac{2}{4\mu^2  + \bar\theta\cdot s} s^{\mu\lambda}\theta_{\lambda\alpha} \, . \label{MP17}
\end{equation}
This coincides with  $T^\mu{}_\nu$ of (\ref{sl0-3}) (taking $s^{\mu\nu}=\frac 12 S^{\mu\nu}$),  then,  (\ref{MP15}) becomes  (\ref{sl0-7}).

\subsection{ Constants of motion.}
%
Depending on the symmetries of the spacetime  studied, the metric could admit  Killing
vectors, $\xi^\mu$ is a Killing vector if $\xi_{\mu;\nu} + \xi_{\nu;\mu}=0$,  such vectors generate constants of motion. For  a spinning particle, the constant of motion associated with $\xi^\mu$  is
\begin{equation}\label{Killing-1}
J^{(\xi)}=P^\mu\xi_\mu - \frac{1}{4}S^{\mu\nu}\nabla_\nu\xi_\mu \, .
\end{equation}
For the case of $\kappa=0$ \cite{Hojman:1978wk, Obukhov:2010kn} can be consulted. For $\kappa \ne 0$, see \cite{Ramirez:2017pmp}.

\section{Spinning particles in Schwarzschild-like spacetimes}\label{S-spacetime}
%
Let us consider the metric in spherical coordinates

\begin{equation}
ds^2  = -f(r) c^2dt^2 + \frac{1}{f(r)} dr^2 +r^2d\theta^2 + r^2\sin^2\theta d\phi^2 
\label{metricfr}\, ,
\end{equation}

Examples of this type of metrics in classical general relativity are Schwarzschild and Reissner-Nordstrom spacetimes. In the context of regular BH, (\ref{metricfr}) includes Ayon-Beato-Garcia (ABG) and Bardeen BHs which have been proved to be exact solutions to the Einstein field equations associated with nonlinear electrodynamics.
The general results found here will be applied to these four working examples.

The set of  phase-space variables which describes the spinning particle is conformed by its position coordinates $x^\mu=(x^0, r, \theta, \phi)$, with canonical momenta 
$P_\mu=(P_0, P_r, P_\theta, P_\phi)$ and the components of the antisymmetric tensor of spin, 
$(S^{0r}, S^{0\phi}, S^{0\theta}, S^{r\theta}, S^{r\phi}, S^{\theta \phi})$.  The Hamiltonian equations  -in a covariant form- are (\ref{motion-P-1})-(\ref{sl4}) for a gravimagnetic particle and (\ref{motion-P-0})-(\ref{sl0-7})  for a MPTD particle. Besides, we have five algebraical  equations; the conditions (\ref{condition2}) and (\ref{mass-shell}), and three independent equations derived from the SSC (\ref{condition1})
\begin{eqnarray}\label{condition1-a}
S^{0i}= \frac{S^{ij} P_j}{P_0} \, .
\end{eqnarray} 
By construction of the vector model, the above mentioned algebraical equations are functionally independent, so they can be used to write five phase-space variables through the others. 
 
Since the metric is rotationally-invariant, it is sufficient to consider the motion on the equatorial plane. Then, the problem is to show that the Hamiltonian equations, together with the algebraic equations (\ref{condition2}), (\ref{mass-shell}) and (\ref{condition1-a}) admit solutions with 
$\theta(\sigma)=\frac \pi 2$ for all $\sigma$. First,  let us establish the conditions for such motion.  If we assume that 
\begin{eqnarray}\label{c1}
P_\theta=0, \quad S^{\theta r}=S^{\theta \phi}=0 , \quad \forall \ \sigma \, ,
\end{eqnarray}
the relations (\ref{condition1-a}) imply 
\begin{eqnarray}
S^{\theta 0} &=& 0 \, , \label{c2}\\
P_0 S^{0r} &=& S^{r\phi} P_\phi \, , \label{S-r0} \\
P_0 S^{0\phi} &=& -S^{r\phi} P_r \, ,\label{S-phi0} 
\end{eqnarray}
through the motion. 
On the other hand, from  (\ref{motion-x-k})  we have that
\begin{eqnarray}
\dot \theta &=& \lambda  \left[P^\theta -\bar a(\kappa -1) S^{\theta \mu}\theta_{\mu\beta} P^\beta \right]   \nonumber \\
&& + \frac{\lambda \kappa \bar a}{8} S^{\theta \mu}(\nabla_\mu  R_{\alpha\beta\sigma\lambda}) S^{\alpha\beta}S^{\sigma\lambda} \, .
\label{motion-x-theta}  
\end{eqnarray}
Then, the conditions (\ref{c1}) and their consequence (\ref{c2}), imply that the angular velocity $\dot\theta$ vanish for all $\sigma$.  It is not difficult to verify that such conditions  are consistent with the equations of motion, i.e.,  $\dot P_\theta=0$ and $\dot S^{\theta \mu}=0$ through the motion. This affirmation is valid for any value of the gravimagnetic moment $\kappa$.

As a result, the phase-space on the equatorial plane is conformed by three coordinates $x^\mu=(x^0, r, \phi)$ their canonical momenta $(P_0, P_r, P_\phi)$ and three non-vanish  components of the spin $(S^{0r}, S^{0\phi}, S^{r\phi})$ (with their antisymmetric partners).
 As a next step, let us show that the dynamic equations for these variables have a solution.
 As we mentioned, we have some independent algebraical  equations,  on the equatorial plane remain four,  (\ref{condition2}),  (\ref{mass-shell}),  (\ref{S-r0}), and (\ref{S-phi0}). Each of these equations can be used to omit one dynamical equation.  In addition to these,  we have two more independent equations  coming from the symmetries of the Schwarzschild-like spacetime. The metric (\ref{metricfr}) admits  four Killing vectors
\begin{eqnarray}
\xi_{(0)}^\mu = (1,0,0,0), \ \xi_{(1)}^\mu = (0,0,\sin{\phi}, -\cot{\theta} \cos{\phi} ),  \\
\xi_{(2)}^\mu= (0,0, \cos{\phi}, -\cot{\theta}\sin{\phi}) , \ \xi_{(3)}= (0,0,0,1)\, .
\end{eqnarray}
For each vector, there is one constant of motion given by (\ref{Killing-1}). On the equatorial plane,
the constants of motion associated with $\xi_{(0)}^\mu$ and $\xi_{(3)}^\mu$ are 
\begin{eqnarray}
J^{(\xi_{0})} = -E&=&P_0 + \frac 14 f'(r) S^{0r}   \, , \label{energy} \\
J^{(\xi_{3})}= L&=&P_\phi + \frac{r}{2} S^{r\phi} \, . \label{ang-m}
\end{eqnarray}
Where $f'(r)=\frac{df(r)}{dr}$. The constant $E$ is the energy of the particle and $L$  its angular momentum. As a result, with (\ref{energy}) and (\ref{ang-m}), we have  six algebraical independent relations among the phase-space coordinates\footnote{ The equations associated to $\xi_{(1)}$ and $\xi_{(2)}$ are not considered since they are fulfilled identically on the equatorial plane.}. 
On the equatorial plane the scalar  
\begin{equation}
\frac{\theta \cdot S}{4} = R_{0r0r} (S^{0r})^2 + R_{0\phi 0\phi} (S^{0\phi})^2 + R_{r\phi r\phi} (S^{r\phi})^2 \, , \label{Theta-S} 
\end{equation}
where 
\begin{eqnarray}
R_{0r0r} &=& \frac 12 f''(r) , \  R_{0\phi 0\phi} = \frac{r}{2} f(r)f'(r) ,  \nonumber \\
R_{r \phi r \phi } &=& -\frac{r}{2} \frac{f'(r)}{f(r)} \, . 
\label{R-comp}
\end{eqnarray}
For the square of the momentum and spin we have
\begin{eqnarray}
P^2&=& -\frac{1}{f(r)} P_0^2 + f(r)P^2_r +\frac{1}{r^2} P^2_\phi \label{PP2} \, , \\
S^2 &=& -(S^{0r})^2 - r^2 f(r) (S^{0\phi})^2 + \frac{r^2}{f(r)}(S^{r\phi}) \, , \nonumber \\
 &=& \left ( \frac{S^{r \phi}}{P_0} \right )^2 \left [ \frac{r^2}{f(r)} P_0^2-r^2 f(r) P_r^2 -P_{\phi}^2 \right] = 4 \alpha
\label{spin-v}
\end{eqnarray}
Using (\ref{S-r0}) and (\ref{S-phi0}) to eliminate the dependence of $S^{r0}$ and $S^{\phi 0}$ in the Killing relations (\ref{energy}) and (\ref{ang-m}) and in conditions  (\ref{condition2}) and (\ref{mass-shell}), we obtain the system 
\begin{eqnarray}
&& P_0\left(P_0 + E \right) + \frac 14 P_\phi S^{r\phi} f'(r)= 0 \, , \label{f1} \\
&& P_\phi + \frac{r}{2} S^{r\phi} - L =0 \label{f2} \, , \\
&&P^2_0 \left[P^2+ \mu^2\right] + \nonumber \\
&& \kappa
\left ( \frac{S^{r\phi}}{2} \right )^2\left[ R_{0r0r} P_{\phi}^2 + R_{0\phi 0\phi} P_r^2 + R_{r\phi r\phi} P_0^2 \right]  =0  \label{f3} \\ 
&&\frac{4\alpha}{r^2} P^2_0 +  (S^{r\phi})^2 P^2 = 0 \, . \label{f4}
\end{eqnarray}
%
%
Solving (\ref{f1})-(\ref{f4}), we can determine  $(P_0, P_r, P_\phi)$ and the component of the spin $S^{r\phi}$  in terms of the radial coordinate $r$, the parameters that $f(r)$ may have and the constants  $(E, L; \alpha)$, i.e., to find the first integrals of motion. After that, the remaining components of the spin are determined using (\ref{S-r0}) and (\ref{S-phi0}). 
At this point, the analysis has been done for arbitrary  gravimagnetic moment $\kappa$. In the next paragraphs, we will study the relevant cases of the unit and zero gravimagnetic moments. 

\subsection{MPTD particles \texorpdfstring{($\kappa=0$)}{k}}
\begin{figure}[ht]
     \includegraphics[width=\linewidth]{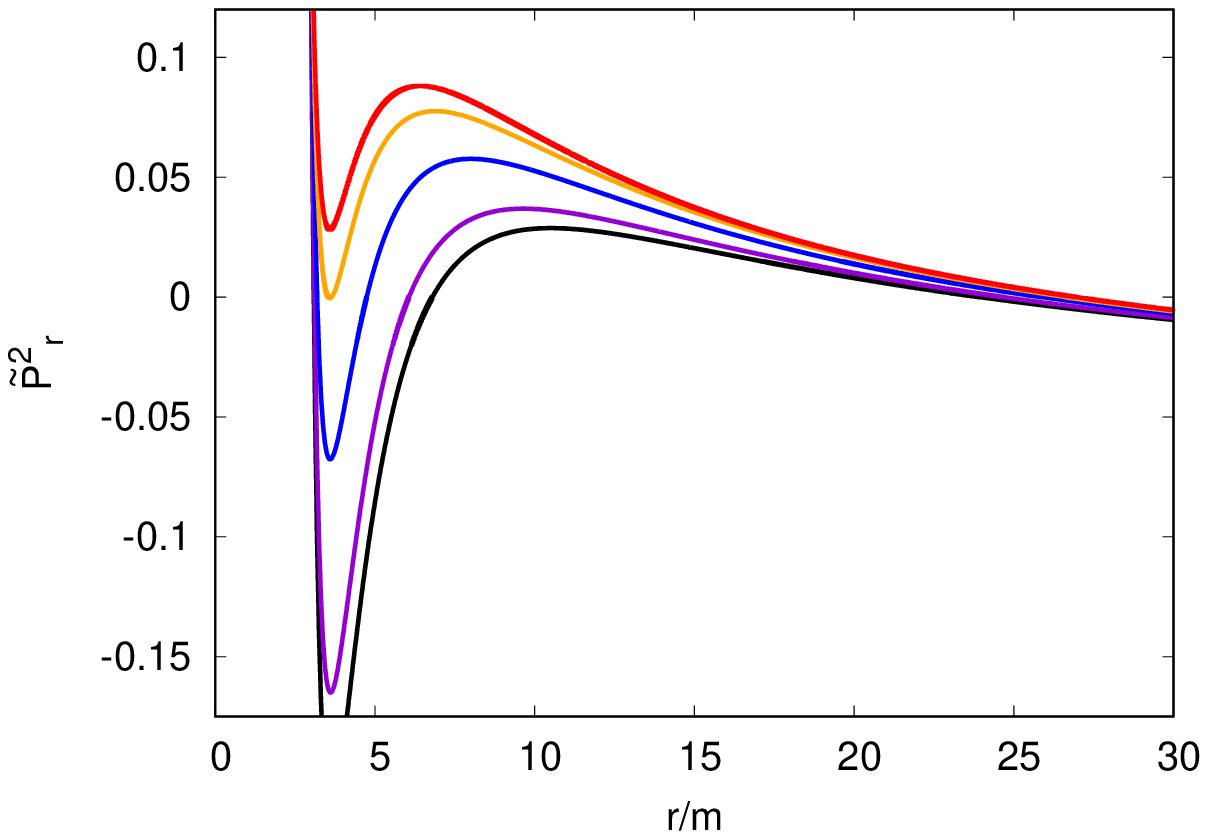}
     \includegraphics[width=\linewidth]{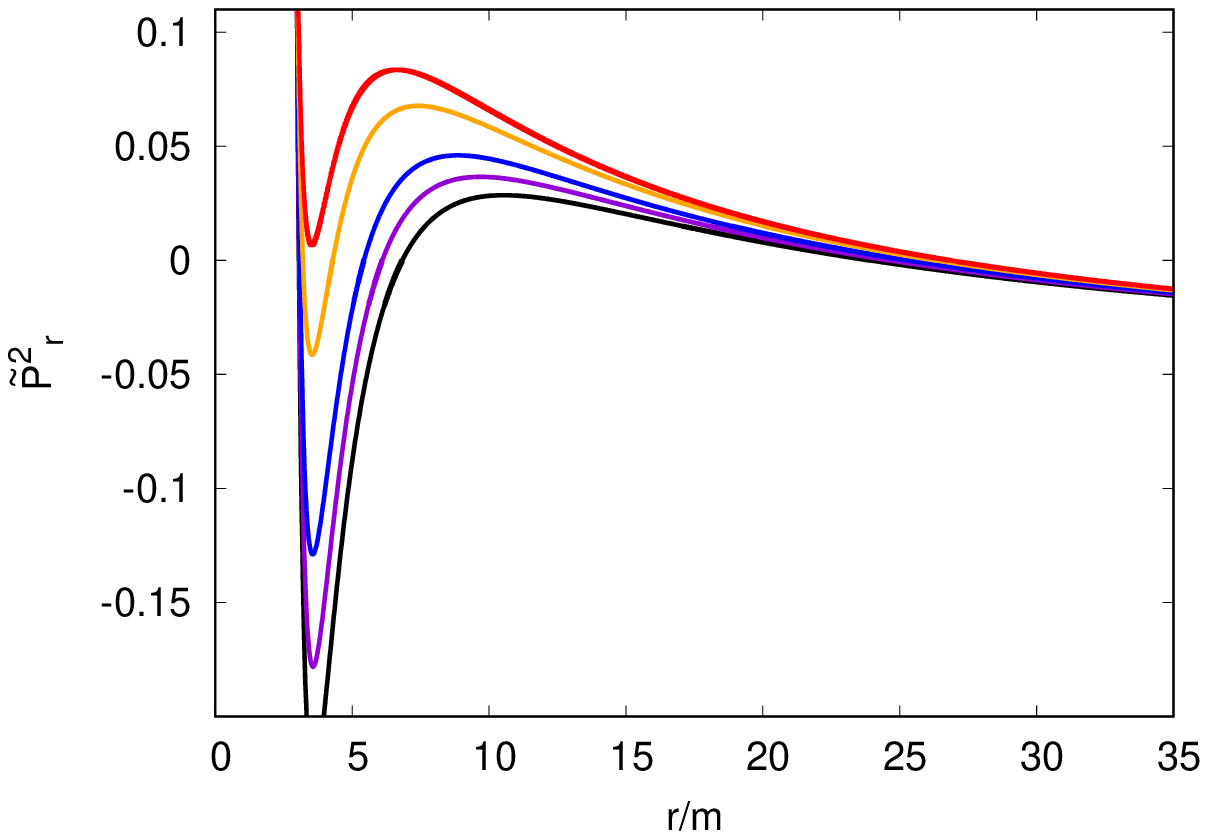}
    \caption{In the upper panel, it is shown $\widetilde{P}_r^2(\widetilde{r})$ with the parameters $\widetilde{E}=0.97$, $\widetilde{L}=4$ and different values of the parameter $\widetilde{a}=0.1$ (the lowest black curve), followed by $\widetilde{a}=0.2, 0.4, 0.54, 0.6$, the last value correspond to the highest red curve, for Schwarzschild BH. In the lower panel, we show $\widetilde{P}_r^2(\widetilde{r})$ for ABG regular BH with $\widetilde{E}=0.97$, $\widetilde{L}=4$ and $\widetilde{a}=0.1$ (the lowest black curve), followed by $\widetilde{a}=0.2, 0.3, 0.5, 0.58$, the last value correspond to the highest red curve.}
    \label{Fig0}
\end{figure}

%
In the case of an MPTD particle, the system (\ref{f1})-(\ref{f4}) has an exact solution. %
For $\kappa=0$, equation (\ref{f3}) becomes (\ref{ms-0}), using this to remove $P^2$ in  (\ref{f4}), we obtain that 
\begin{equation}
\left(S^{r\phi}  - \sqrt{\frac{ 4\alpha }{ \mu^2 r^2 }} P_0  \right)\left( S^{r\phi} + \sqrt{\frac{ 4\alpha }{ \mu^2 r^2 }} P_0  \right)=0 \, , \label{f-34}
\end{equation}
which implies
\begin{equation}
S^{r\phi}= \pm \frac{ 2 a}{ r} P_0 \, . \label{PS-k0}
\end{equation}
where $a^2=\alpha/\mu^2$. On the other hand, combining  (\ref{f2}) and (\ref{f1}), we get  the equation
\begin{equation}
P_0 \left( P_0 + E \right) - \frac{r}{8} f'(r) S^{r\phi} \left( S^{r\phi} - \frac 2r L\right) =0 \, . \label{f-12}
\end{equation}
Combining this with   (\ref{PS-k0})  we obtain 
\begin{equation}
P^{(\pm)}_0 = - \frac { E \pm f'(r) \frac{a L}{2 r}  }{ 1 - \beta  } \, .
\label{Pcero-k0}
\end{equation}
Once $P_0$ has been determined, $S^{r \phi}$ can be found using  (\ref{PS-k0}) and $P_\phi$ using (\ref{f2}), resulting
\begin{eqnarray}
S^{r\phi}_{(\pm)} =&& - \left( \frac 2r \right)  \frac{ \beta L \pm a E   } { 1 - \beta } 
\label{PS-k0-a} \, , \\
P_\phi^{(\pm)} =&&   \frac{  L \pm a E }{  1 - \beta } , \, 
\label{Pphi-k0-a}
\end{eqnarray}
where we have introduced $\beta \equiv a^2 f'(r)/2r$. Finally, replacing  (\ref{Pcero-k0}) and (\ref{Pphi-k0-a}) in (\ref{ms-0}), that is, in $P^2+\mu^2=0$, $P_r$ is readily attained
\begin{eqnarray}
 P_r^2 &=&  \frac{1}{[f(r)]^2} \left( \frac { E \pm f'(r) \frac{ a L}{2 r}}{1- \beta} \right)^2 
 \nonumber \\
 && - \frac{1}{f(r)} \left[ \frac{ 1}{ r^2 }  \left(   \frac{  L \pm a E }{  1 - \beta }   \right)^2 + \mu^2 \right] \, .  
 \label{Pr-k0}
\end{eqnarray}
For $f(r)=1-2m/r$, the Schwarzschild solution for the MPTD equations which have been reported in \cite{Suzuki:1996gm, Hojman:2012me} is recovered. The expressions for $P_0,P_{\phi},P_r$ and $S^{r \phi}$
presented here allows to use space-times of the form (\ref{metricfr}) like the Reissner-Nosrdstrom BH as well as Bardeen and Ayon-Beato-Garcia BG regular BHs among others \cite{Bardeen, Ayon-Garcia, Eloy-Alberto}. In any case, given a set of parameters $E,L,a,\mu$, and the parameters encoded in the specific space-time, the condition $P_r^2>0$ provides the radial spatial domain for the solution. It is convenient to introduce the variables scaled with $\mu$ and $m$ (the mass parameter of the black hole being studied)
\begin{eqnarray}
\widetilde{S}^{r \phi} &\equiv& S^{r \phi}/\mu, \quad \widetilde{S}^{0\phi}\equiv S^{0\phi}/\mu, \quad \widetilde{S}^{0 r}\equiv S^{0 r}/m \mu, \nonumber \\
\widetilde{P}_r &\equiv& P_r/\mu, \quad \widetilde{P}_{\phi}\equiv /m \mu, \quad \widetilde{P}_0\equiv P_0/\mu, \quad \tilde{E} \equiv E/\mu, \nonumber \\ 
\widetilde{L} &\equiv& L/(m\mu), 
\quad \widetilde{r} \equiv r/m , \quad \widetilde{a}\equiv a/m,
\label{Escalamientos}
\end{eqnarray}
\\
Given a set of parameters $\widetilde{a},\widetilde{E},\widetilde{L}$, we still have to decide which sign to use in (\ref{Pcero-k0}). Using the lower sign in (\ref{Pcero-k0}) or equivalently in (\ref{Pr-k0}), we may determine the region in space where $P_r^2>0$ holds. For $\widetilde{E}=0.97$, $\widetilde{L}=4.0$, in the upper panel in figure \ref{Fig0}, we show $\widetilde{P}_r^2(\widetilde{r})$ for Schwarzschild BH with different values of $\widetilde{a}=0.1$ (the lowest black curve), followed by $\widetilde{a}=0.2, 0.4, 0.54, 0.6$, the last value correspond to the highest red curve. For 
$\widetilde{a}<\widetilde{a}_c \approx 0.5337$ there are three roots of $\widetilde{P}_r^2(\widetilde{r})$: $\widetilde{r}_1<\widetilde{r}_2<\widetilde{r}_3$. It turns out that  $\widetilde{P}_r^2(\widetilde{r})$ is
positive for $\widetilde{r}_H<\widetilde{r}<\widetilde{r}_1$ and $\widetilde{r}_2<\widetilde{r}<\widetilde{r}_3$, where the scaled $\widetilde{r}_H\equiv=r_H/m=2$ is the location of Schwarzschild event horizon. As $\widetilde{a}$ climbs up,  $\widetilde{r}_1$ and $\widetilde{r}_2$ get closer until they collide at $\widetilde{a}_c \approx 0.5337$ and disappear for $\widetilde{a}>\widetilde{a}_c$ leaving  $\widetilde{P}_r^2(\widetilde{r})$ with just one root, $\tilde{r}_3$, for $\widetilde{r}<\widetilde{r}_3$, $\widetilde{P}_r^2(\widetilde{r})$ is positive. 

In the second panel in figure \ref{Fig0}, we show $\widetilde{P}_r^2(\widetilde{r})$ for ABG regular BH whose $-g_{tt}=f(r)$ metric component is given by

\begin{equation}
    f(r)= 1-\frac{2 m r^2}{(r^2+q^2)^{3/2}}+\frac{q^2 r^2}{(r^2+q^2)^2}, \nonumber
\end{equation}

\noindent with different values of the parameter $\widetilde{a}=0.1$ (the lowest black curve), followed by $\widetilde{a}=0.2, 0.3, 0.5, 0.58$, the last value correspond to the highest red curve and $\widetilde{q}\equiv q/m=0.1$. In this case $\widetilde{a}_c \approx 0.55804$.
For Reissner-Nordstrom BH, with $\widetilde{E}=0.97$, $\widetilde{L}=4.0$ and $\widetilde{q}=0.1$, the critical values of the parameter $\widetilde{a}$ is $\widetilde{a}_c=0.547320$ and $\widetilde{a}_c=0.54406$ for Bardeen regular BH. In both of these cases, $\widetilde{P}_r^2(\widetilde{r})>0$ in the radial intervals $\widetilde{r}_H^{ext}<\widetilde{r}<\widetilde{r}_1$ and $\widetilde{r}_2<\widetilde{r}<\widetilde{r}_3$, as long as $\widetilde{a}<\widetilde{a}_c$. Reissner-Nordstrom, Bardeen and ABG possess an exterior and an interior event horizons $\widetilde{r}_H^{int}$ and $\widetilde{r}_H^{ext}$ for certain values of q and m, specifically the exterior one for Reissner–Nordstrom  is $\widetilde{r}_H^{ext}=1+\sqrt{1-\widetilde{q}^2}$. For Bardeen regular BH, the event interior and exterior horizons must be determined numerically by solving the equations 

\begin{equation}
\widetilde{r}^6 +(3 \widetilde{q}^2-4) \widetilde{r}^4 +3 \widetilde{q}^4 \widetilde{r}^2 + \widetilde{q}^6=0. \nonumber
\end{equation}
\noindent For ABG regular BH, the event horizon must be found numerically as well by solving a polynomial. At any rate, for both types of space times, we work in the interval $\widetilde{r}>\widetilde{r}_H^{ext}$. These three roots $\widetilde{r}_1<\widetilde{r}_2<\widetilde{r}_3$ are the turning points, defined as those points where the radial momentum $\widetilde{P}_r$ vanishes. 

\begin{figure}[ht]
     \includegraphics[width=\linewidth]{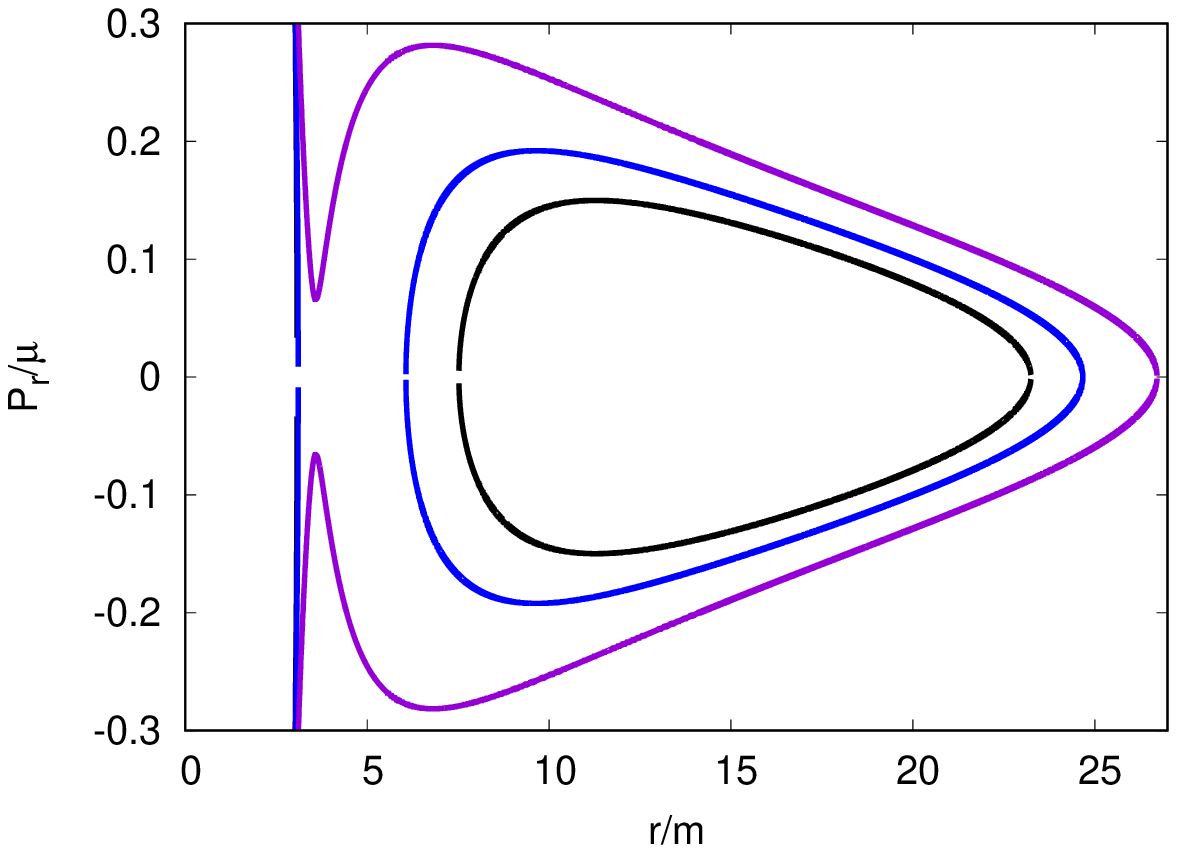}
     \includegraphics[width=\linewidth]{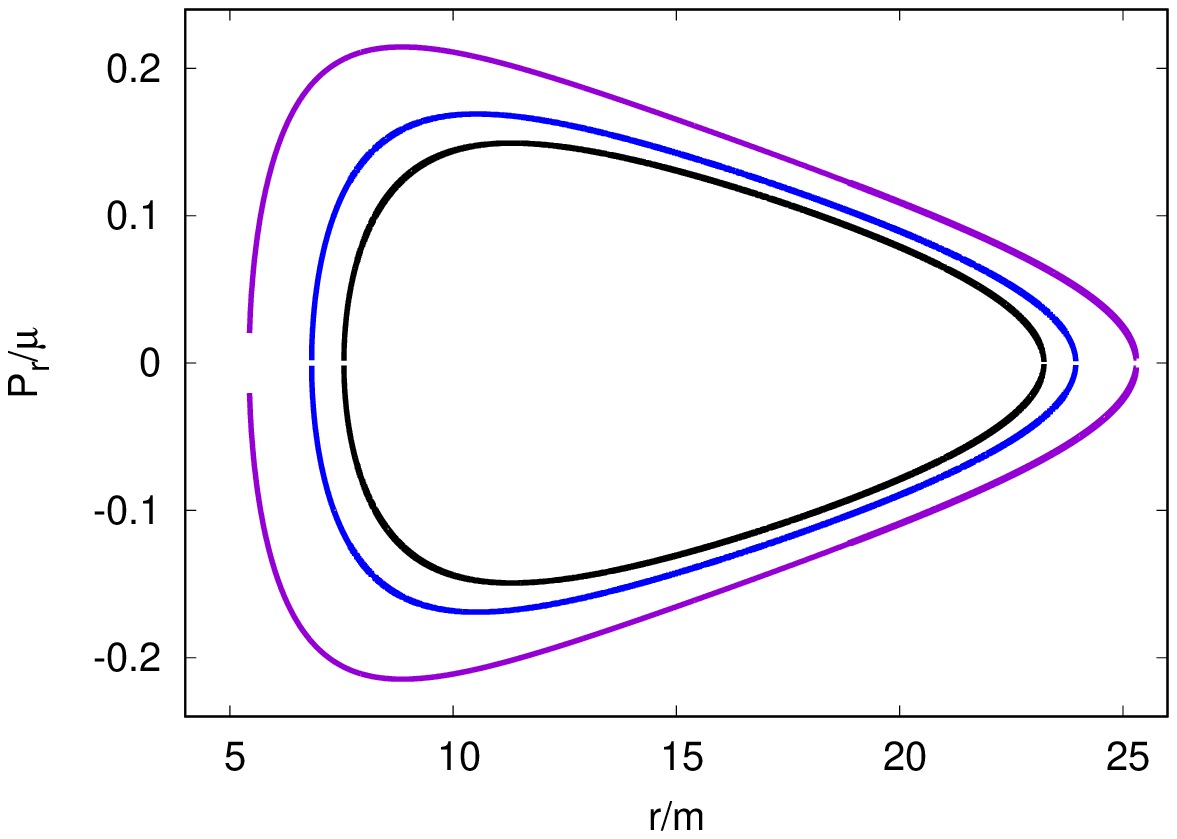}
    \caption{Radial momentum $P_r/\mu\equiv \widetilde{P}_r$ for energy and angular momentum $\widetilde{E}=0.97$, $\widetilde{L}=4.0$ of spinning particles around a Schwarzschild BH is shown in the first plot for $\widetilde{a}=0.01$, black curve which consist of two parts, one open curve for $\widetilde{r}_H=2<\widetilde{r}<\widetilde{r}_1=3.07475$ and a close cycle for $\widetilde{r}_2=7.52084<\widetilde{r}<\widetilde{r}_3=23.2453$. For $\widetilde{a}=0.2$ one has also two parts (blue curve) one open and one close curve. For $\widetilde{a}=0.55$ the turning points $\widetilde{r}_1$ and $\widetilde{r}_2$ have disappeared and the corresponding curve in violet colour is an open one. In the lower plot with $\widetilde{E}=0.97$, $\widetilde{L}=4.0$, we show only the cycle section of $\widetilde{P}_r(\widetilde{r})$ for the Ayon-Beato-Garcia regular BH for $\widetilde{q}=0.1$ and $\widetilde{a}=0.01,0.1,0.3$ (black, blue and violet curves respectively).
    }
    \label{Fig1}
\end{figure}

The upper plot in figure \ref{Fig1} shows the radial momentum $\widetilde{P}_r(\tilde{r})$ of spinning particles around the Schwarzschild black hole, we use the following values for energy and angular momentum parameters $\widetilde{E}=0.97$, $\widetilde{L}=4.0$ and different values of the spinning parameter  $\widetilde{a}=0.01,0.2,0.55$. For the first two values, there are three turning points and the graph of $\widetilde{P}_r(\widetilde{r})$ has two sections, one for $\widetilde{r}_H=2<\widetilde{r}<\widetilde{r}_1$ which is an open trajectory in phase space and the other for $\widetilde{r}_2<\widetilde{r}<\widetilde{r}_3$ which is a close cycle. For $\widetilde{a}=0.55>\widetilde{a}_c$ there is solely one turning point; thus a particle may travel up to that point and then returns. In the lower plot of figure \ref{Fig1}, with $\widetilde{E}=0.97$, $\widetilde{L}=4.0$ as well, we show only the cycle section of $\widetilde{P}_r(\widetilde{r})$ for the Ayon-Beato-Garcia regular BH with $\widetilde{q}=0.1$ and $\widetilde{a}=0.01,0.1,0.3$ (black, blue and violet curves respectively). We observed that the area of each cycle is enlarged for both, Schwarzschild and ABG BHs, as $\widetilde{a}$ increases. Exactly the same behavior is exhibited by Reissner-Nordstrom and Bardeen BHs. \\

Some turning points are shown in table \ref{tabla1} for $\widetilde{E}=0.97$, $\widetilde{L}=4.0$ and for different values of the parameter $\widetilde{a}$ for Schwarzschild, Reissner-Nordstrom BHs as well as for Bardeen and ABG regular BHs. In all cases, as $\widetilde{a}$ increases, the first turning point $r_1$ increases whereas the second one $r_2$ decreases and the third one $\widetilde{r}_3$ increases its value. 
The first two turning points $\widetilde{r}_1$ and $\widetilde{r}_2$ collapse and disappear at $\widetilde{a}_c$, leaving just one turning point $\widetilde{r}_3$. The critical parameters are $\widetilde{a}_c=0.5337, 0.54732, 0.55804, 0.54406$ for Schwarzschild, Reissner–Nordstrom , ABG and Bardeen BHs respectively. \\

For Reissner–Nordstrom  BHs as well as for Bardeen and ABG space-times, we vary the charge parameter $\widetilde{q}$, keeping $\widetilde{E}=0.97$, $\widetilde{L}=4.0$, and show in table \ref{tabla2} the corresponding turning points $\widetilde{r}_2=\widetilde{r}_{min}$ and  $\widetilde{r}_3=\widetilde{r}_{max}$ for the cycle sections. It is observed that as $\widetilde{q}$ increases, the lower turning point $\widetilde{r}_{min}$ undergoes an increment, yet the upper turning point $\widetilde{r}_{max}$ decreases its value in all cases. It is for Bardeen regular BH that $\widetilde{r}_{max}$ streaches farther away. For $\widetilde{q}<0.769$ and for $\widetilde{q}<0.6342$ the Barden and ABG space-times are regular BHs respectively, otherwise they are globally regular space-times \cite{Bardeen} - \cite{Eloy-Alberto}.
\begin{widetext}

\begin{table}[ht]
    \centering
    \begin{tabular}{|c|rrr|rrr|rrr|rrr|}
    \hline 
    Parameter & \multicolumn{3}{|c|}{Schwarzschild} & \multicolumn{3}{|c|}{Reissner-Nordstrom} & \multicolumn{3}{|c|}{Ayon-Beato-Garcia} & \multicolumn{3}{|c|}{Bardeen} \\
    \hline
    \hline
     $\widetilde{a}$ & $\widetilde{r}_1$ & $\widetilde{r}_2$ & $\widetilde{r}_3$ & $\widetilde{r}_1$ & $\widetilde{r}_2$ & $\widetilde{r}_3$  & $\widetilde{r}_1$ & $\widetilde{r}_2$ & $\widetilde{r}_3$ & $\widetilde{r}_1$ & $\widetilde{r}_2$ & $\widetilde{r}_3$ \\
     \hline
     $0.01$ & $3.07475$ & $7.51984$ & $23.24021$ & $3.05927$ & $7.54309$ & $23.22742$ & $3.04418$ & $7.55229$ & $23.22577$ & $3.05961$ & $7.52911$ & $23.85600$ \\
    \hline
     $0.1$ & $3.08663$ & $6.79552$ & $23.95264$ & $3.06971$ & $6.81915$ & $23.94102$ & $3.05355$ & $6.82943$ & $23.93957$ & $3.07038$ & $6.80589$ & $23.95119$ \\
    \hline
     $0.2$ & $3.10607$ & $6.06352$ & $24.66466$ & $3.08690$ & $6.08861$ & $24.65403$ & $3.06907$ & $6.10066$ & $24.65273$ & $3.08808$ & $6.07575$ & $24.66337$ \\
  \hline
     $0.3$ & $3.13684$ & $5.38642$ & $25.31028$ & $3.11407$ & $5.41454$ & $25.30040$ & $3.09361$ & $5.42945$ & $25.29924$ & $3.11605$ & $5.40167$ & $25.30911$ \\
    \hline
    \hline
    \end{tabular}
    \caption{Values for the turning points $\widetilde{r}_1<\widetilde{r}_2<\widetilde{r}_3$ of MPTD ($\kappa=0$) spinning particles orbiting in our four space-time working examples with $\widetilde{E}=0.97$, $\widetilde{L}=4.0$. In general, as $\widetilde{a}$ increases, $r_1$ increases as well, $\widetilde{r}_2$ decreases and $\widetilde{r}_3$ increases. The charge value was chosen as $\widetilde{q}=0.1$. For $\widetilde{a}>\widetilde{a}_c$ the first two turning points disappears leaving only one turning point, a particle may travel up to that unique point and then return. The critical parameters are $\widetilde{a}_c=0.5337, 0.54732, 0.55804, 0.54406$ for Schwarzschild, Reissner-Nordstrom, ABG and Bardeen BHs respectively.}
    \label{tabla1}
\end{table}

\begin{table}[ht]
    \centering
    \begin{tabular}{|rr|rr|rr|rr|}
    \hline 
    \multicolumn{2}{|c|}{Parameters} & \multicolumn{2}{|c|}{Reissner-Nordstrom} & \multicolumn{2}{|c|}{Ayon-Beato-Garcia} & \multicolumn{2}{|c|}{Bardeen} \\
    \hline
    \hline
     $\tilde{a}$ & $\widetilde{q}$ & $\tilde{r}_2$ & $\tilde{r}_3$ & $\tilde{r}_2$ & $\tilde{r}_3$  & $\tilde{r}_2$ & $\tilde{r}_3$ \\
     \hline
     \multirow{3}{*}{$0.01$} & $0.3$ & $7.7270$  & $23.1299$ & $7.8040$ & $23.1148$ & $7.6025$ & $23.2309$ \\
    & $0.6$ & $8.3165$ & $22.7698$ & $8.5735$ & $22.7050$ & $7.8320$ & $23.1861$ \\
     & $0.9$ & $9.2590$ & $22.1201$ & $9.7490$ & $21.9535$ & $8.1725$ & $23.1105$ \\
  \hline
     \multirow{3}{*}{$0.1$} & $0.3$ & $7.0040$  & $23.8529$ & $7.0880$ & $23.8396$ & $6.8750$ & $23.9450$ \\
    & $0.6$ & $7.5815$ & $23.5281$ & $7.8475$ & $23.4719$ & $7.1355$ & $23.9055$ \\
     & $0.9$ & $8.4695$ & $22.9509$ & $8.9520$ & $22.8105$ & $7.4925$ & $23.8391$ \\
     \hline
     \multirow{3}{*}{$0.3$} & $0.3$ & $5.6250$  & $25.2262$ & $5.7350$ & $25.2156$ & $5.5165$ & $23.3050$ \\
    & $0.6$ & $6.2215$ & $24.9528$ & $6.5215$ & $24.9086$ & $5.8350$ & $25.2733$ \\
     & $0.9$ & $7.0515$ & $24.4760$ & $7.5425$ & $24.3693$ & $6.2445$ & $25.2202$ \\
    \hline
    \hline
    \end{tabular}
    \caption{Values for the smallest $\widetilde{r}_2$ and largest $\widetilde{r}_3$ turning points in the close curves section of MPTD ($\kappa=0$) spinning particles orbiting around a Reissner-Nordstrom BH as well as ABG and Bardeen regular space-times with $\widetilde{E}=0.97$, $\widetilde{L}=4.0$ fixed and for some parameter sets $(\widetilde{a},\widetilde{q})$. We observe that as $\widetilde{q}$ increases, $\widetilde{r}_2$ also increases yet $\widetilde{r}_3$ decreases in all cases. It is for Bardeen regular BH that $\widetilde{r}_{max}$ is stretches farther away.}
    \label{tabla2}
\end{table}

\end{widetext}
Had we chosen the upper sign in (\ref{Pr-k0}), or equivalently $P_0^{(+)}$ in (\ref{Pcero-k0})
\begin{equation}
P^{(\pm)}_0 = - \frac { E \pm f'(r) \frac{a L}{2 r}  }{ 1 - \beta  } \nonumber \, .
\end{equation}

\noindent what we would have observed is that, the area of the $(\widetilde{r},\widetilde{P}_r)$ cycle instead of undergoing an increment as the one shown in figure \ref{Fig1}, it would have shrunken as $\tilde{a}$ increases until it reaches a value $a_c$ where the lower and upper turning points collide. 
There are no further solutions beyond this value $a_c$ and there is no physical reason for this bound. This feature is in contrast to the lower sign choice in (\ref{Pr-k0}), that means, the choice of $P_0^{(-)}$, which does not present this rare behaviour. In the literature the lower sign is chosen (see for instance \cite{Hojman:2012me}), basically because working with the upper sign would correspond to 4-momentum pointing toward the past \cite{Misner}. Hence, we will choose the lower sign.

\subsection{Gravimagnetic particle \texorpdfstring{($\kappa=1$)}{k}}
We are now going to analyse whether there is a solution for the momenta $P_{\nu}(r)$ and $S^{r \phi}(r)$ for a given set of parameters $E,L,\mu,\alpha$ of the system (\ref{f1})-(\ref{f4}) when $\kappa=1$, the so called gravimagnetic particles. It is convenient to rewrite the system in an apposite form. It turns out that the variable $P_0$ totally decouples. Three components of the Riemann tensor $R_{0r0r}, R_{0\phi 0 \phi}$ and $R_{r \phi r \phi}$ are needed for the gravimagnetic case. All Riemann tensor components are provided at the appendix.  After some algebraic steps, it is possible to transform the set of equations (\ref{f1})-(\ref{f4}) into the equivalent system
 \begin{equation}
 P_\phi = - \frac{r}{2} S^{r\phi} + L , 
 \label{g2}
 \end{equation}
 
 \begin{equation}
 -P^2 = \mathcal B(P_0), \label{g1} 
 \end{equation}
 
 \begin{equation}
 (S^{r\phi})^2\mathcal B(P_0) =\frac{4\alpha}{r^2} P_0^2 , \label{g4} 
 \end{equation}
 
 \begin{equation}
 S^{r\phi}\left( S^{r\phi} - \frac{2}{r} L \right) = P_0 \mathcal A(P_0) , \label{g3}
 \end{equation}
 where we have introduced $\mathcal B(P_0)$, a second-order polynomial on $P_0$, defined as
 \begin{eqnarray}
 \mathcal B(P_0) &\equiv &  \mu^2 -\frac{\alpha f'(r)}{2r} - 2 \frac{ (P_0 + E)^2 }{(f'(r))^2}\left[ \frac{f'(r)}{r} -f''(r) \right]  \, , \nonumber \\
 &=& \mu^2 \left ( 1-\beta (r) - \frac{2 a_f(r)}{(f'(r))^2} \left [ \frac{P_0+E}{\mu} \right ]^2 \right ) ,
 \label{BP0}
 \end{eqnarray}
 here we have employed the previously introduced function 
 \begin{equation}
 \beta (r) \equiv \frac{a^2 f'(r)}{2r},
 \end{equation}
 recall that $a^2 \equiv \alpha/\mu^2$. Since $\alpha$ represents the square of the internal angular momentum of the test body, $a$ is its angular momentum per unit mass with dimensions of length, just like in the Kerr solution for a stationary body with angular momentum. The function $a_f(r)$ that appears in (\ref{BP0}), is defined as
 \begin{equation}
 a_f(r) \equiv \frac{f'(r)}{r} - f''(r)
 \end{equation}
 The first-order polynomial on $P_0$ appearing in (\ref{g3}) $\mathcal A(P_0)$, is defined as
 \begin{equation}
 \mathcal A(P_0)= \frac{ 8 \left( P_0 + E \right) }{r f'(r)}  \, .
 \end{equation}
 
 We see that  equations (\ref{g4}) and (\ref{g3}) contain only $P_0$, $S^{r\phi}$ and $r$. Our goal now is to prove that this subsystem has a solution, that is to say, that is possible to find $P_0$ and $S^{r\phi}$ as functions of $r$ for a given set of parameters $\{ E,L,\mu,\alpha \}$.  Combining (\ref{g4}) and (\ref{g3}) the following equivalent subsystem is readily attained 
 
\begin{eqnarray}
&& (S^{r\phi})^2\mathcal B(P_0) = \frac{4\alpha}{r^2} P_0^2 \, \label{g4bis} \\
&& \left [ Q(P_0) \right ]^2=\frac{16 L^2 \alpha}{r^4} \mathcal B(P_0) \, . \label{g3bis} 
\end{eqnarray}

Notice that (\ref{g3bis}) is an equation only for the momentum $P_0$, as mentioned before, the variable $P_0(r)$ has been decoupled from the other momenta and $S^{r \phi}$. It turns out to be that (\ref{g3bis}) is a six order polynomial in the momentum $P_0(r)$. Hence, in order to find $P_0(r)$, one has to find the roots of this polynomial for each $r$ in a radial domain where these roots actually exist, this task will have be performed numerically. 

In (\ref{g3bis}), $Q(P_0)$ is a cubic polynomial for $P_0$, whose definition involves the linear function $\mathcal A(P_0)$ and the quadratic function $\mathcal B(P_0)$ in the following manner

\begin{widetext}
\begin{equation}
  Q(P_0) =  \mathcal A(P_0) \mathcal B(P_0) - \frac{4 \alpha P_0}{r^2}
    = -\frac{16\mu^3 a_f(r)}{r \left ( f'(r) \right )^3} \left ( \left ( \frac{P_0+E}{\mu}\right)^3 - \frac{(f'(r))^2}{2a_f(r)} \left [ (1-2\beta(r)) \frac{P_0+E}{\mu}+\beta \frac{E}{\mu}\right ] \right )
     \label{QPP0}
\end{equation}
\end{widetext}
 
 For material particles $P^2<0$, as a result (\ref{g1}) implies that the quadratic function $\mathcal B(P_0)>0$. The sign of $a_f(r)$ determines whether the parabola $\mathcal B(P_0)$ has a maximum or a minimum. The roots of $\mathcal B(P_0)$ are given by
 
 \begin{equation}
     P_{\mathcal B 0}^{(\pm)}= - E \pm \mu \sqrt{\left [ 1-\beta(r)\right ] \frac{(f'(r))^2}{2 a_f(r)} }
     \label{raicesB}
 \end{equation}
 
From (\ref{BP0}) we observe that if $a_f(r)>0$ then $\mathcal B(P_0)$ has a maximum and it will be positive solely in the interval between the roots (\ref{raicesB}) which are real provided that $\beta(r) < 1$. Therefore, we must find solutions of the sixth order polynomial (\ref{g3bis}) written now as
 
\begin{equation}
 \mathcal P_6(P_0)= \left [ Q(P_0) \right]^2 -\frac{16 L^2 \alpha}{r^4} \mathcal B(P_0) 
 \label{P6}
\end{equation}
exclusively in the interval $I_{\mathcal B} = (P_{\mathcal B 0}^{-},P_{\mathcal B 0}^{+})$ in order that (\ref{g1}) makes sense for material particles. Since $Q(P_0)$ is a cubic equation, it has always at least one real root. Lets denote such a root by 
$P_0^{*}$, if $P_0^{*} \in I_{\mathcal B}$ and $a_f(r)>0$, then $\mathcal P_6(P^{*}_0)<0$ since 
$\mathcal B(P_0)>0$. Due to the fact that $\mathcal P_6(P_0) \to \infty$ as $P_0 \to \pm \infty$, we conclude that $\mathcal P_6(P_0)$ has at least two real roots. At any rate, to have a consistent system, at least, one root of $\mathcal P_6(P_0)$ should be in the interval $I_{\mathcal B}$ to guarantee existence of solutions for the system, if that is the case, the remaining variables are computed using (\ref{g2}), (\ref{g1}) and (\ref{g4}). To prove analytically that, given a set of parameters $E,L,a,\mu$, at least one root of  $\mathcal P_6(P_0)$ is located in the interval $I_{\mathcal B}$ may be utterly cumbersome and difficult. Instead, we will provide numerical proof of its existence. It is convenient to use the scaled variables (\ref{Escalamientos}), thereby, the system (\ref{g2})-(\ref{g3}) is rewritten with only three parameters $\widetilde{a},\widetilde{E},\widetilde{L})$ instead of five as follows
\begin{eqnarray}
\widetilde{P}_{\phi} &=& -\frac{\widetilde{r}}{2}\widetilde{S}^{r \phi}+\widetilde{L} \label{UNO} \\
\widetilde{P}_r^2 &=& \frac{\widetilde{P}_0^2}{f^2(\widetilde{r})} - \frac{1}{f(\widetilde{r})}\left [ \frac{\widetilde{P}^2_{\phi}}{\widetilde{r}^2} + \widetilde{\mathcal{B}}(\widetilde{P}_0) \right ] \label{DOS}\\
\left ( \widetilde{S}^{r\phi} \right )^2 &=& \frac{4 \widetilde{a}^2 \widetilde{P}_0^2}{\widetilde{r}^2 \widetilde{\mathcal{B}}(\widetilde{P}_0)} \label{TRES} \\
\mathcal{P}_6(\widetilde{P}_0) &=& \left [ \widetilde{Q} (\widetilde{P}_0) \right ]^2 - 
\frac{m^2[f'(\tilde{r})]^6\widetilde{L}^2 \widetilde{a}^2}{16 a_f^2(\tilde{r})\widetilde{r}^2} \widetilde{\mathcal{B}}( \widetilde{P}_0)=0 . \nonumber \\
\label{CUATRO}
\end{eqnarray}
This is our basic system we will be working with from now on. The quadratic function $\widetilde{\mathcal{B}}(\widetilde{P}_0)$ reads 
\begin{equation}
\widetilde{\mathcal{B}}(\widetilde{P}_0) = 1-\beta (\widetilde{r}) - \frac{2 a_f(\widetilde{r})}{(f'(\widetilde{r}))^2} \left [ \widetilde{P}_0 + \widetilde{E} \right ]^2 
\label{Bescalada} 
\end{equation}
and the cubic function $\widetilde{Q}(\widetilde{P}_0)$ is now written as
\begin{eqnarray}
\widetilde{Q}(\widetilde{P}_0) &=& \left ( \widetilde{P}_0 + \widetilde{E} \right)^3 \nonumber \\
&-&\frac{(f'(\widetilde{r}))^2}{2a_f(\widetilde{r})} \left [ (1-2\beta(\widetilde{r})) (\widetilde{P}_0+\widetilde{E}) + \beta(\widetilde{r}) \widetilde{E} \right ] \nonumber \\
&=& \widetilde{P}_0^3+A\widetilde{P}_0^2+B\widetilde{P}_0+C
\end{eqnarray}
whose coefficients are explicitly given as
\begin{eqnarray}
    A&=& 3 \widetilde{E} \nonumber \\
    B&=& 3\widetilde{E}^2 - \frac{(f'(\widetilde{r}))^2}{2 a_f(\widetilde{r})} \left[ 1-2 \beta (\widetilde{r}) \right] 
    \nonumber \\
    C&=& \widetilde{E}^3 - \frac{(f'(\widetilde{r}))^2}{2 a_f(\widetilde{r})} \left[ 1- \beta (\widetilde{r}) \right] \widetilde{E}
    \label{CoeficientesGenerales}
\end{eqnarray}
The coefficients of the six degree polynomial $P_6(\widetilde{P}_0) = \sum_{n=0}^6 K_n \widetilde{P}_0^n$ are given by
\begin{eqnarray}
K_0 &=& C^2 + \mathcal{K} \left ( \widetilde{E}^2-(1-\beta(\tilde{r}))\frac{[f'(\tilde{r})]^2}{2a_f(\tilde{r})} \right ), \nonumber \\
K_1 &=&  2 B C + 2\mathcal{K} \widetilde{E}, \nonumber \\
K_2 &=& B^2+2AC+\mathcal{K}, \nonumber \\
K_3 &=& 2(A B+C), \nonumber \\
K_4 &=& A^2 + 2 B, \nonumber \\
K_5 &=& 6 \widetilde{E} \quad \text{and} \quad K_6=1, 
\label{CoefsP6}
\end{eqnarray}
with $A,B$ and $C$ are given in (\ref{CoeficientesGenerales}) and
\begin{equation}
    \mathcal{K}=\frac{m^2 \widetilde{L}^2 \widetilde{a}^2 [f'(\tilde{r})]^4}{8\tilde{r}^2 a_f(\tilde{r})} \quad  \text{and} \quad \beta(\tilde{r})=\frac{\widetilde{a}^2f'(\tilde{r}) m}{2 \tilde{r}}
\end{equation}

We have achieved our goal: $P_0$ have been decoupled, to find it, we only need to find roots of the polynomial $\mathcal{P}_6$ given a set of parameters $\widetilde{a},\widetilde{E},\widetilde{L}$, yet keeping only those roots which lie on the interval $I_{\mathcal{B}}$. Now we proceed to numerically solve the system for gravimagnetic particles for the Schwarzschild and Reissner-Nosrdström BHs as well as for the Bardeen and ABG regular BHs. These last two spacetimes are exact solutions of Einstein field equation with nonlinear electromagnetism \cite{Bardeen, Ayon-Garcia, Eloy-Alberto}. \\  

The corresponding $f(r)$, $a_f(r)$ and $\beta(r)$ for these four spacetimes are given next.\\

\noindent {\bf Schwarzschild}
\begin{equation}
    f(r)= 1-\frac{2m}{r} , \quad a_f(r)= \frac{6m}{r^3} , \quad \beta(r)= \frac{a^2 m}{r^3}
    \label{trioSchwardzschild}
\end{equation}
\noindent {\bf Reissner-Nordstrom}
\begin{eqnarray}
f(r) &=& 1-\frac{2m}{r}+\frac{q^2}{r^2} , \quad a_f(r)= \frac{6 m r-8q^2}{r^4}, \nonumber \\
\beta(r) &=& \frac{a^2(m r-q^2)}{r^4}
\label{trioRN}
\end{eqnarray}
\noindent {\bf Bardeen}
\begin{eqnarray}
f(r) &=& 1- \frac{2mr^2}{R^{3/2}} , \quad a_f(r)=6m r^2\frac{r^2-4q^2}{R^{7/2}}, 
\nonumber \\
\beta(r) &=& a^2 m\frac{r^2-2q^2}{R^{5/2}}
\label{trioBardeen}
\end{eqnarray}
\noindent {\bf Ayon-Beato-Garcia}
\begin{eqnarray}
f(r) &=& 1-\frac{2 m r^2}{R^{3/2}}+\frac{q^2 r^2}{R^2}, \nonumber \\
a_f(r) &=& 2 r^2 \frac{4q^2(2q^2-r^2) \sqrt{R}+3m(r^4-3q^2r^2-4q^4)}{R^{9/2}}
\nonumber \\
\beta(r) &=& a^2 \frac{q^2(q^2-r^2)\sqrt{R}+m(r^4-r^2 q^2-2q^4)}{R^{7/2}}
\label{trioABG}
\end{eqnarray}
where $R \equiv r^2+q^2$. In all cases, as $q \to 0$ we recover $f(r),a_f(r)$ and $\beta(r)$ for Schwarzschild BH. \\

\begin{figure}[ht]
    \centerline{\epsfysize=3cm \epsfbox{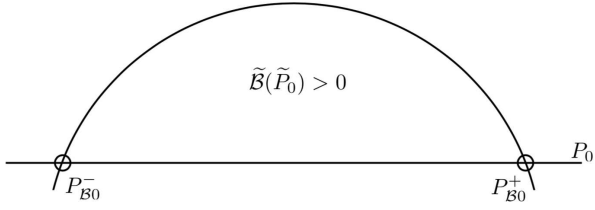}}
    \caption{Due to $a_f>0$ for Schwarzschild BH, the parabola $\mathcal{B}(P_0)$ is positive in the interval $I_{\mathcal{B}}=({P}^{-}_{\mathcal{B} 0},{P}^{+}_{\mathcal{B} 0})$. To have a solution physically acceptable, at least one real root of the six order polynomial $P_6(P_0)$ must be located in this interval. See text for details.}
    \label{Parabola}
\end{figure}

We work first with the Scwarzschild BH, in this case $a_f(r)=6m/r^3$ which is always positive; therefore $\mathcal{B}(r)$ is a downward parabola whose roots are given as
\begin{equation}
    \widetilde{P}^{\pm}_{\mathcal{B} 0}= -\widetilde{E} \pm \sqrt{\frac{1-\beta}{3\widetilde{r}}}=-\widetilde{E}\pm \sqrt{\frac{\widetilde{r}^3-\widetilde{a}^2}{3 \widetilde{r}^4}},
    \label{P0raiceSch}
\end{equation}
which are real quantities provided that $1-\beta>0$, or equivalently $\widetilde{r}>\widetilde{a}^{2/3}$. In order to get $\widetilde{P}_0(\widetilde{r})$, one has to find the roots of the polynomial $\mathcal{P}_6(\widetilde{P}_0)$ given in(\ref{CUATRO}). As mentioned before, the system (\ref{UNO})-(\ref{CUATRO}) has a consistent solution provided that at least one root of $\mathcal{P}_6(\widetilde{P}_0)$ is located in the interval where $\widetilde{\mathcal{B}}(\widetilde{P}_0) > 0$. Since $a_f>0$ for Schwarzschild BH, we have a downwards parabola, and this interval is $I_{\mathcal{B}}=(\widetilde{P}^{-}_{\mathcal{B} 0},\widetilde{P}^{+}_{\mathcal{B} 0})$ (see figure \ref{Parabola}). 

Given a set of fixed parameters $\widetilde{a},\widetilde{E},\widetilde{L}$, we vary $\widetilde{r}>\widetilde{a}^{2/3}$ and at each value of $\widetilde{r}$, we search for roots of the polynomial $\widetilde{P}_6(\widetilde{P}_0)$ only in the interval $I_{\mathcal{B}}$.
Any of the Newton's methods family can be employed, we shall use a hybrid method of bisection and Newton-Raphson \cite{nr}. We summarize the numerical algorithm now
%

\begin{enumerate}
    \item An arbitrary fixed set of parameter $\widetilde{a}, \widetilde{E}, \widetilde{L}$ is first given together with an arbitrary radial domain $I_{r}=[\widetilde{r}_{min},\widetilde{r}_{max}]$ with $\widetilde{r}_{min}> \widetilde{r}^{ext}_H$, being $\widetilde{r}^{ext}_H$ the (exterior) event horizon.
    \item We construct a partition of $I_r$ as $\widetilde{r}_n= \widetilde{r}_{min}+ n \Delta_N$ from
    $n=0,...,N$ with $\Delta_N= (\widetilde{r}_{max} - \widetilde{r}_{min})/N$.
    \item Then we search for real roots of $\mathcal{P}_6(\widetilde{P}_0(\widetilde{r}_n))=0$ solely in the interval $I_{\mathcal B} = (P_{\mathcal B 0}^{-},P_{\mathcal B 0}^{+})$ where $\mathcal{B}(\widetilde{P}_0)>0$. If such real roots exist in that interval, we call them $P_0^{\star}(\widetilde{r}) \in I_{\mathcal B}$.
    \item For each real root $P_0^{\star}(\widetilde{r}_n)$, $\widetilde{S}^{r \phi}(\widetilde{P}^{\star}_0(\widetilde{r}_n))$ is computed using (\ref{TRES}) and $\widetilde{P}_{\phi}(\widetilde{P}_0^{\star}(\widetilde{r}_n))$ is computed with (\ref{UNO}).
    \item $\widetilde{P}^2_r(\widetilde{P}_0^{\star}(\widetilde{r}_n))$ is calculated using (\ref{DOS}), if this quantity is positive, this data set is stored to plot the momenta (and $\widetilde{S}^{r \phi}(\widetilde{P}_0(\widetilde{r}_n))$), otherwise is discarded. 

    \item We select the next radial coordinate $\widetilde{r}_{n+1}$ and continue until we reach $\widetilde{r}_{max}$.
    
\end{enumerate}

{\bf Gravimagnetic solutions for $L=0$}\\

According to (\ref{CUATRO}), in order to find $\widetilde{P}_0(\widetilde{r})$ for $L=0$, one deals with the cubic  $\widetilde{Q}(\widetilde{P}_0(\widetilde{r}))=0$ instead of the six order polynomial $\widetilde{P}_6(\widetilde{P}_0(\widetilde{r}))=0$. A cubic equation might have either one real and two complex roots or three real roots all at once. For gravimagnetic particles, any root $\widetilde{P}_0^{*}$ must imply $\mathcal{B}(\widetilde{P}_0^{*})>0$ to be physically meaningful. In all the surveys performed, following the algorithm just described above, we found three real roots, yet only two of them fulfilled this requirement.\\

We show in table \ref{tabla3} some examples of the parameter space $\widetilde{a},\widetilde{E},\widetilde{L}=0$ where these such two solutions of the system (\ref{UNO})-(\ref{CUATRO}) indeed exist. $\widetilde{r}_{min}$ is chosen arbitrarily a little greater than $\widetilde{r}^{ext}_H= 2$, the Schwarzschild BH event horizon. 
For a given set of parameters $\widetilde{a},\widetilde{E},\widetilde{L}=0$, $\widetilde{r}_{tp}$ represents the spinning particle's turning point, defined as the point where the radial component of the momentum $P_r$ vanishes. In other words, solutions of our system (\ref{UNO})-(\ref{CUATRO}) for gravimagnetic particles indeed exist for 
$\widetilde{r} \le \widetilde{r}_{tp}^{(i)}$ with $i=1,2$. Although mathematically there are two solutions fulfilling the requirement $\mathcal{B}(\widetilde{P}_0^{*})>0$, thereby two 
$\widetilde{r}_{tp}^{(i)}$ turning points for each single set of parameters $\tilde{a},\tilde{E},\tilde{L}=0$, only $\widetilde{r}_{tp}^{(1)}$ is admissible since 
\begin{equation}
    \lim_{\widetilde{r} \to \widetilde{r}_{tp}} \widetilde{P}_0 = -\widetilde{E} \nonumber
\end{equation}
holds solely for $\widetilde{r}_{tp}^{(1)}$ not for $r_{tp}^{(2)}$. This assertion was numerically verified, for instance, for Schwarzschild BH with $\tilde{a}=0.01 , \tilde{E}=0.8$, as $\widetilde{r} \to \widetilde{r}_{tp}^{(2)}$, the momentum $\widetilde{P}_0 \to -1.02576$ while as $\widetilde{r} \to \widetilde{r}_{tp}^{(1)}$ $\widetilde{P}_0 \to -0.8000065$. As a result, we keep only the $\widetilde{P}_r$ branch whose corresponding turning point is $\widetilde{r}_{tp}^{(1)}$ and disregard the other branch. Hereafter by $\widetilde{r}_{tp}$ we mean $\widetilde{r}_{tp}^{(1)}$. A spinning particle may travel away from the BH up until $\widetilde{r}_{tp}$ where its radial momentum $\widetilde{P}_r$ vanishes, and then return.\\

 The second column shows $r_{tp}$ for Schwarzschild BH. Other examples of parameter sets $\widetilde{a},\widetilde{E},\widetilde{L}=0$ where solutions exist, together with the corresponding turning points, are constructed and shown in the same table \ref{tabla3}, for Reissner-Nordstrom BH, ABG and Bardeen regular black holes (with $\widetilde{q}=0.5$) as well. In all the surveys we performed, we observed that, for a fixed value of $\tilde{a}$, as $\tilde{E}$ increases, the turning point $\tilde{r}_{tp}$ increases as well (not linearly). In all cases, as $\tilde{E}$ is near unity, $\tilde{r}_{tp} \to \infty$. The further away finite turning point shown in this table corresponds to Schwarzschild BH followed by Bardeen, then Reissner-Nordstrom and ABG BHs. As $\tilde{a}$ increases (and for a fixed $\widetilde{E}$ close to unity), the turning points values (slightly) decrease their values for the four space-times working examples considered here.\\

\begin{figure}
     \includegraphics[width=\linewidth]{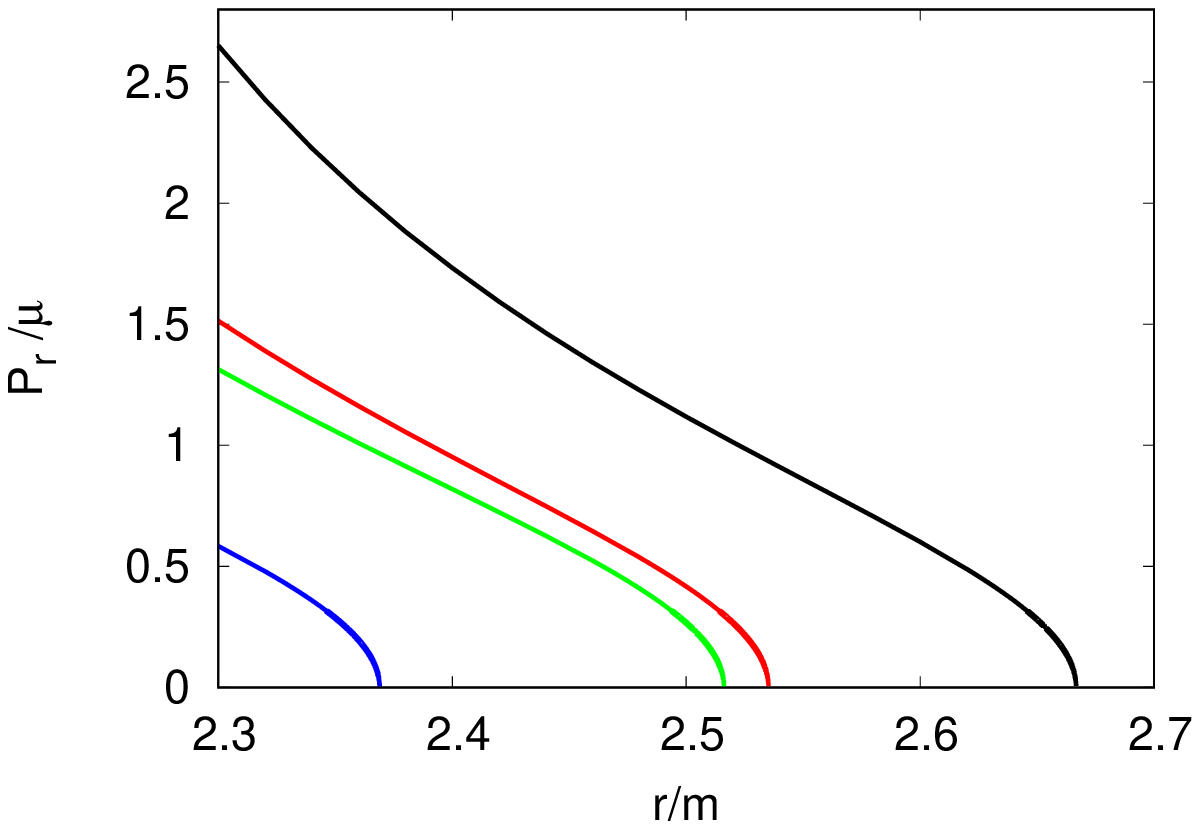}
     \includegraphics[width=\linewidth]{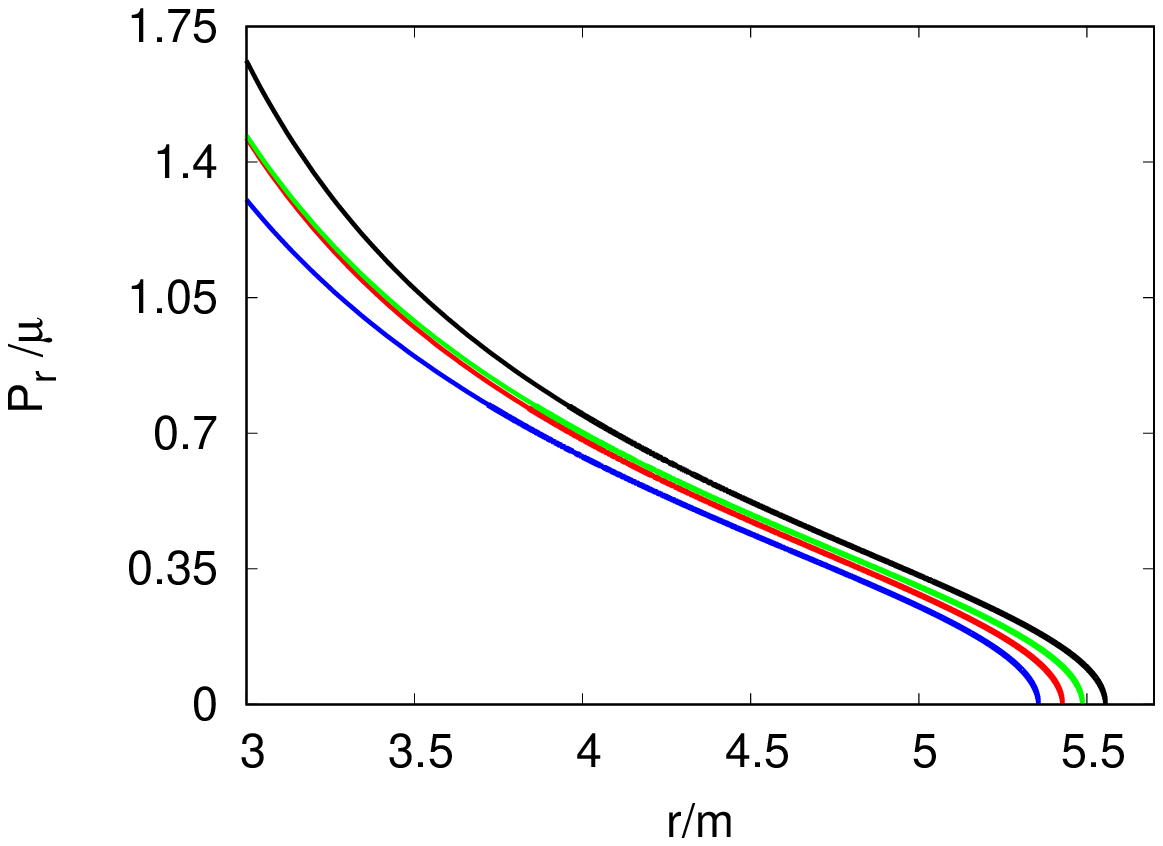}
    \caption{We show $\widetilde{P}_{r}(\widetilde{r})$ for  $\widetilde{a}=0.01, \widetilde{L}=0$ and $\widetilde{E}=0.5$ (upper panel) as well as $\widetilde{E}=0.8$ (lower panel) for gravimagnetic particles. These curves are numerically generated following the algorithm described in the text, and correspond to the first and second row of table \ref{tabla3} respectively. The upper (black curves) correspond to Schwarzschild BH, the second (red) curves to Reissner-Nordstrom, the third (green) curves to ABG and the fourth and lowest (blue) curves to Bardeen regular BH.}
    \label{PrTP}
\end{figure}

Plots of $\widetilde{P}_{r}(\widetilde{r})$ for  $\widetilde{a}=0.01, \widetilde{L}=0$ and $\widetilde{E}=0.5,0.8$ are shown in figure \ref{PrTP}. The curves in the upper panel correspond 
to the very first row in table \ref{tabla3}. The upper (black) curve represents $\widetilde{P}_r(\tilde{r})$ that crosses the horizontal axis at $\tilde{r}_{max}=2.6667$ 
for Schwarzschild BH. The second curve (in red)
corresponds to the radial momentum that vanishes at $\tilde{r}_{max}=2.5332$ for Reissner-Nordstrom BH. The third (green) curve is $\widetilde{P}_r(\tilde{r})$ that vanishes at $\tilde{r}_{max}=2.5162$ and the fourth lower curve (in blue) corresponds to the radial momentum for ABG BH with $\tilde{r}_{max}=2.3689$. The curves in the lower panel in figure \ref{PrTP} correspond to the second row in table \ref{tabla3} for our four spacetime working examples.\\

\begin{widetext}

\begin{table}[ht]
    \centering
    \begin{tabular}{|rr|rr|rr|rr|rr|}
    \hline 
    \multicolumn{2}{|c|}{Parameters} & \multicolumn{2}{|c|}{Schwarzschild} & \multicolumn{2}{|c|}{Reissner-Nordstrom} & \multicolumn{2}{|c|}{Ayon-Beato-Garcia} & \multicolumn{2}{|c|}{Bardeen} \\
    \hline
    \hline
     $\tilde{a}$ & $\widetilde{E}$  & $\tilde{r}^{(1)}_{tp}$ & $\tilde{r}^{(2)}_{tp}$ & $\tilde{r}^{(1)}_{tp}$  & $\tilde{r}^{(2)}_{tp}$ & $\tilde{r}^{(1)}_{tp}$  & $\tilde{r}^{(2)}_{tp}$ & $\tilde{r}^{(1)}_{tp}$  & $\tilde{r}^{(2)}_{tp}$ \\
     \hline
     \multirow{4}{*}{$0.01$} & $0.50$ & $2.6667$ & $4.9211$ & $2.5332$  & $4.7043$ & $2.3689$ & $4.4827$ &  $2.5162$ & $4.7209$ \\
    & $0.80$ & $5.5555$ & $6.5448$ & $5.4276$ & $6.3075$ & $5.3562$ & $6.1197$ & $5.4870$ & $6.3717$ \\
     & $0.90$ & $10.5262$ & $7.1751$ & $10.3997$ & $6.9323$ & $10.3632$ & $6.7553$ & $10.4904$ & $7.0116$\\
    \hline
     \multirow{4}{*}{$0.1$} & $0.50$ & $2.6678$ & $4.9206$ & $2.5363$ & $4.7038$ & $2.3701$ & $4.4821$ & $2.5173$ & $4.7204$ \\
    & $0.80$ & $5.5552$ & $6.5446$ & $5.4272$ & $6.3073$ & $5.3558$ & $6.1195$ & $5.4866$ & $6.3715$ \\
     & $0.90$ & $10.5241$ & $7.1750$ & $10.3976$ & $6.6322$ & $10.3610$ & $6.7552$ & $10.4882$ & $7.0115$\\
    \hline
     \multirow{4}{*}{$0.3$} & $0.50$ & $2.6765$ & $4.9164$ & $2.5454$ & $46994$ & $2.3793$ & $4.4775$ & $2.5262$ & $4.7159$ \\
    & $0.80$ & $5.5526$ & $6.5432$ & $5.4242$ & $6.3058$ & $5.3522$ & $6.1179$ & $5.4836$ & $6.3700$ \\
     & $0.90$ & $10.5070$ & $7.1526$ & $10.3800$ & $6.6314$ & $10.3432$ & $6.7544$ & $10.4710$ & $7.0107$ \\
  \hline
     \multirow{4}{*}{$0.5$} & $0.50$  & $2.6942$  & $4.9080$ & $2.5637$ & $4.6906$ & $2.3980$ & $4.4683$ & $2.5443$ & $4.7071$\\
    & $0.80$ & $5.5477$ & $6.5402$ & $5.4182$ & $6.3027$ & $5.3454$ & $6.1148$ & $5.4778$ & $6.3669$ \\
     & $0.90$ & $10.4730$ & $7.1725$ & $10.3448$ & $6.6297$ & $10.3076$ & $6.7527$ & $10.4366$ & $7.0091$\\
    \hline
    \hline
    \end{tabular}
    \caption{We show some examples of parameter sets $\tilde{a},\tilde{E},\tilde{L}=0$ for which solutions of the system for gravimagnetic spinning particles exist. There are two solution branches that end up at $r_{tp}^{(i)}$ ($i=1,2$) which are turning points, defined as the points where the radial component of the momentum $P_r$ vanishes. It turns out that only the branch related to $r_{tp}^{(1)}$ is acceptable (see text for details). For Reissner-Nordstrom, Bardeen and ABG BHs, the charge parameter was chosen as $\tilde{q}=0.5$}
    \label{tabla3}
\end{table}

\end{widetext}

{\bf Gravimagnetic solutions for $L \neq 0$}\\

\begin{figure}
\includegraphics[width=\linewidth]{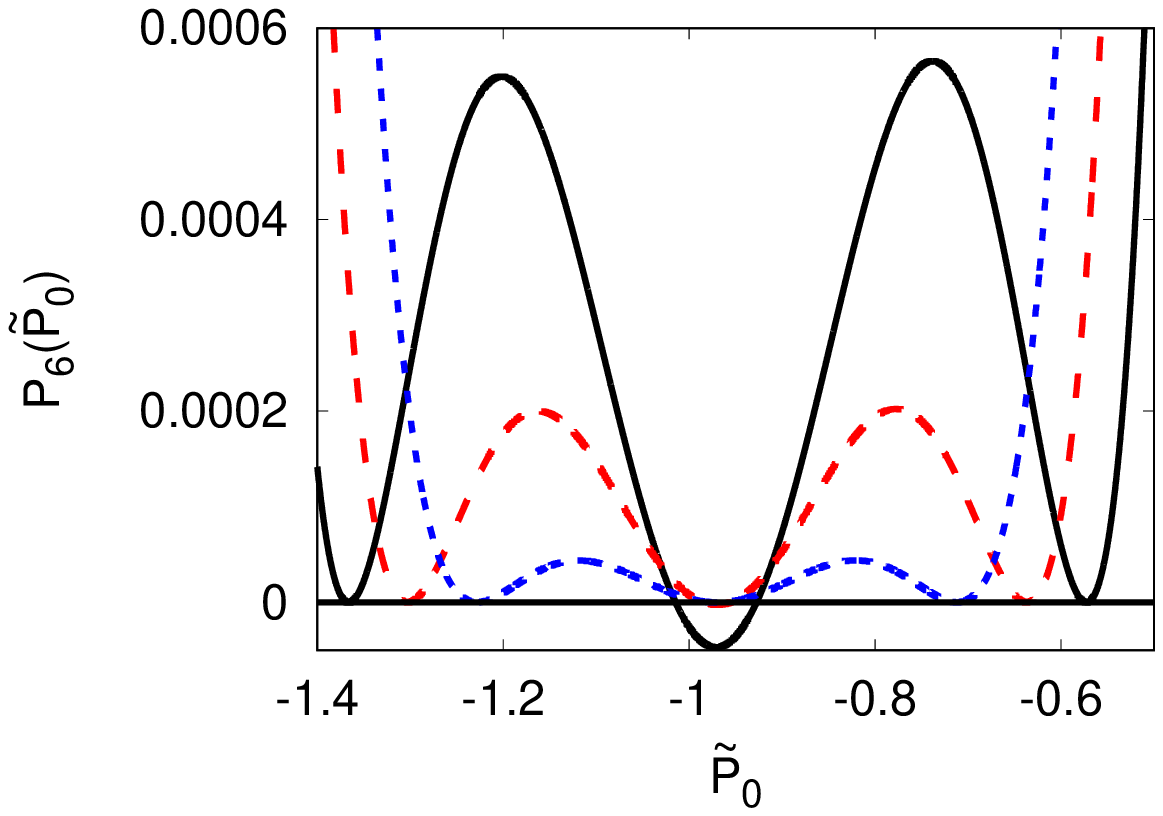}
    \begin{tabular}{cc}
\resizebox{42mm}{36mm}{\includegraphics{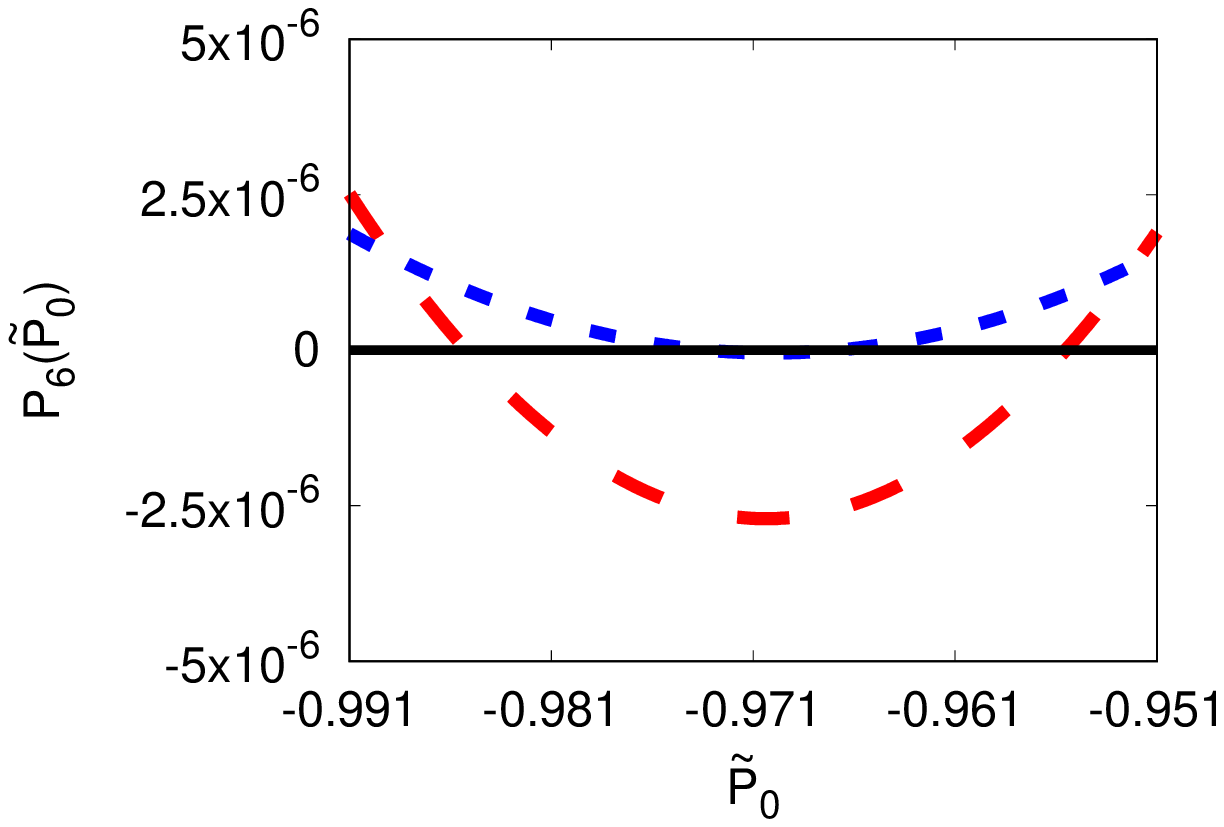}} &
\resizebox{42mm}{36mm}{\includegraphics{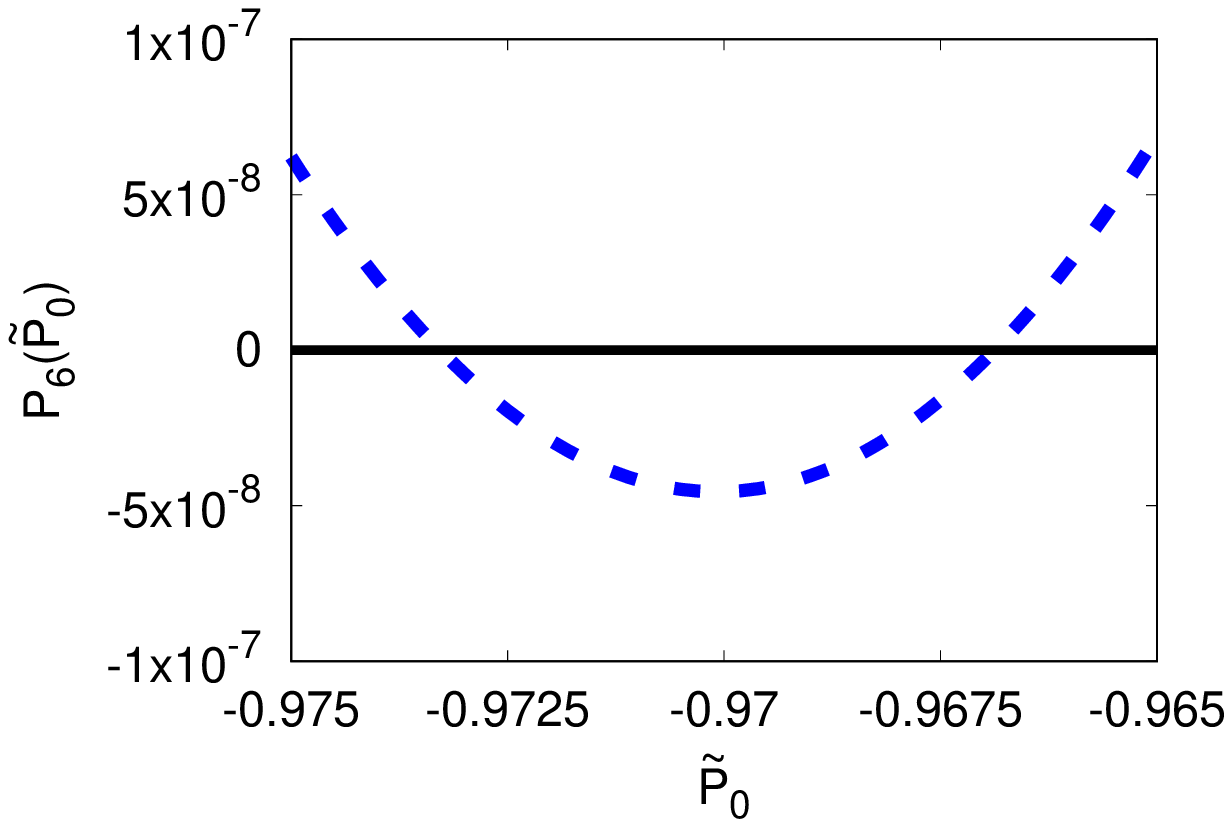}} \\
\end{tabular}
    \caption{In the upper panel, it is shown, for the Schwarzschild case, the polynomial $\mathcal{P}_6(\widetilde{P}_0(\widetilde{r}))$ for $\widetilde{r}=2.1$ (solid black line), $\widetilde{r}=3.0$ (dash red line), $\widetilde{r}=5.0$ (dotted blue line) and the parameters $\widetilde{a}=0.1, \widetilde{E}=0.97, \widetilde{L}=4.0$. In the lower graphs, the sixth degree polynomial is enhanced for $\widetilde{r}=3.0,5.0$ to make evident that indeed, there are roots in the vicinity of $\widetilde{P}_0= -\widetilde{E}=- 0.97$. }
 \label{GraficaP6}
\end{figure}

We shall now be concerned with the non zero angular momentum case. Given a set of parameters $\widetilde{a},\widetilde{E},\widetilde{L}$, by following our numerical algorithm described in this section, we construct solutions for our system (\ref{UNO})-(\ref{CUATRO}) for gravimagnetic spinning particles. The momenta $\widetilde{P}_0(\widetilde{r})$ is now obtained by finding the roots of the sixth degree polynomial $\mathcal{P}_6(\widetilde{P}_0)$, in principle, for at any value of the radial coordinate $r$, up to six roots might be found. The question about what roots we should work with again arise. One may notice from the polynomial

\begin{equation}
\mathcal{P}_6(\widetilde{P}_0) = \left [ \widetilde{Q} (\widetilde{P}_0) \right ]^2 - 
\frac{m^2[f'(\tilde{r})]^6\widetilde{L}^2 \widetilde{a}^2}{16 a_f^2(\tilde{r})\widetilde{r}^2} \widetilde{\mathcal{B}}( \widetilde{P}_0)\nonumber
\end{equation}
that it is the second term on the right hand side, that could make $\mathcal{P}_6(\widetilde{P}_0)$ have negative values, thereby $\mathcal{P}_6(\widetilde{P}_0)$ could have roots. Figure \ref{GraficaP6} shows the $\mathcal{P}_6(\widetilde{P}_0(\widetilde{r}))$ for different values of $\widetilde{r}$ and 
$\widetilde{a}=0.1, \widetilde{E}=0.97, \widetilde{L}=4.0$ for the Schwarzschild case. Regions on the $\widetilde{P}_0$-axis where $\mathcal{P}_6(\widetilde{P}_0)<0$ are exceedingly narrow; in the top graph in figure \ref{GraficaP6}, one can clearly observe that region at first glance just for $\widetilde{r}=2.1$ (solid black curve) and in the vicinity of $\widetilde{P}_0=-0.97$. For $\widetilde{r}=3.0,5.0$, enhanced graphs of the polynomial around $\widetilde{P}_0=-0.97$ are shown, it is then apparent that $\mathcal{P}_6(\widetilde{P}_0)<0$; hence roots exist there. One also notices that all roots approaches $-\widetilde{E}$ as $\widetilde{r}$ increases, as it should be. We employed quadruple precision in the roots finding hybrid numerical method since the intervals where $\mathcal{P}_6(\widetilde{P}_0)<0$ are tiny, moreover, for instance, for $\widetilde{r}=5.0$, one observes (see figure \ref{GraficaP6}) that 
$-10^{-8} < \mathcal{P}_6(\widetilde{P}_0) < 0$; with just single o double precision, a root may have been missed.\\

\begin{figure}[ht]
     \includegraphics[width=\linewidth]{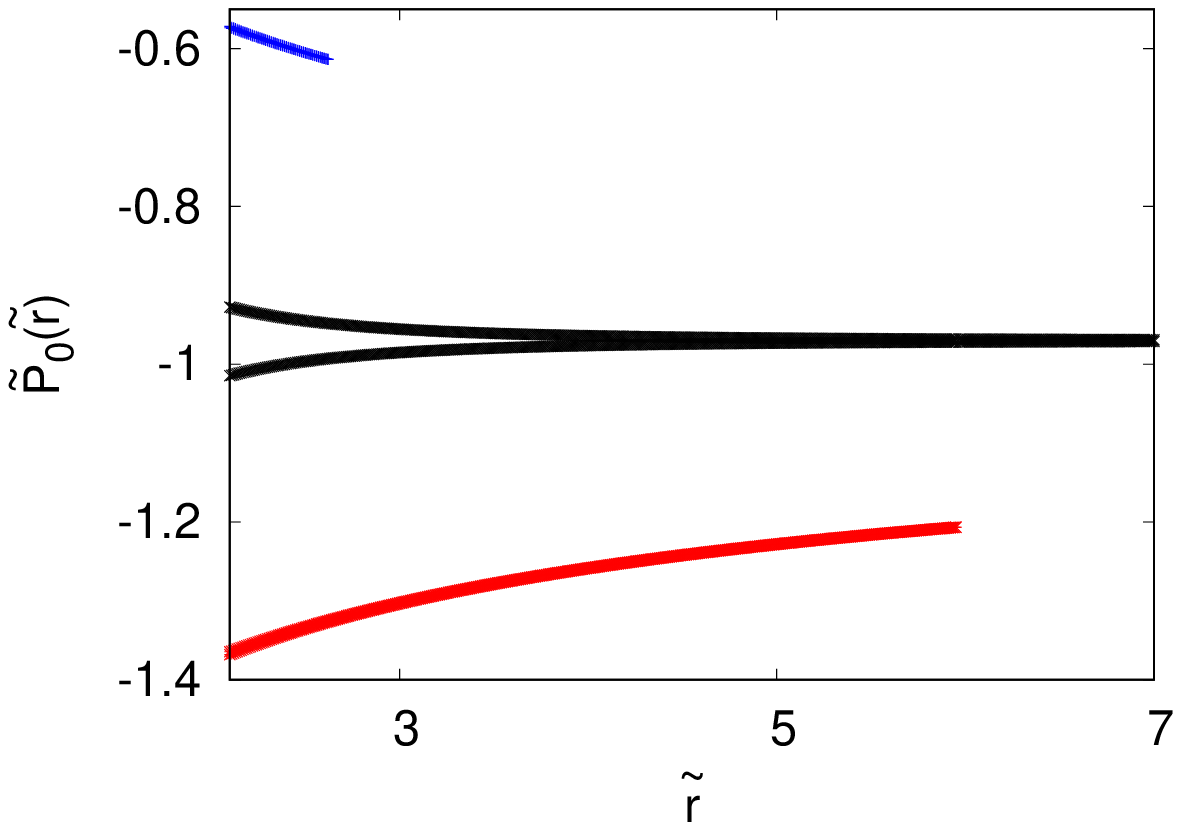}
    \caption{For the Schwarzschild BH case, we show the six roots of the polynomial $\mathcal{P}_6(\widetilde{P}_0(\widetilde{r}))$ for $\widetilde{a}=0.1, \widetilde{E}= 0.97, \widetilde{L}=4.0$, for this set of parameters, the roots come in pairs. There are two roots in the lower red curve which ends at $\widetilde{r}=5.96$, there are another two roots in the upper blue curve which ceases to exist at $\widetilde{r}=2.63$. The roots in the middle (black curves) remain all along our numerical domain, they rapidly approach $-\widetilde{E}$, one from above $\widetilde{P}^{(a)}_0 \equiv \widetilde{P}^{\star}_0> -\widetilde{E}$, and the other from below $\widetilde{P}^{(b)}_0 \equiv \widetilde{P}^{\star}_0<-\widetilde{E}$.   }
    \label{SeisRaices}
\end{figure}

For small values of $\widetilde{r}$, usually six roots are found, they come in rather close pairs, two of these pairs cease to exist as $\widetilde{r}$ increases. Only the pair of roots near to $-\widetilde{E}$ remains, and this pair satisfy $\widetilde{P}_0(\widetilde{r}) \to -\widetilde{E}$ as $\widetilde{r} \to \infty$. This behavior is shown in figure \ref{SeisRaices} for the Schwarzschild case, with the parameters $\widetilde{a}=0.1, \widetilde{E}=0.97, \widetilde{L}=4.0$. The upper (blue) pair of roots, which are very close and are indistinguishable by sight, ceases to exist at $\widetilde{r}=2.63$, the lower (red) pair of roots, which are also indistinguishable by sight, terminates at $\widetilde{r}=5.96$, solely the middle pair of roots remains for all values of $\widetilde{r}$ greater than the event horizon. 

Similar plots to those shown in figure \ref{SeisRaices}, exist for our other three working examples, Reissner-Nordstrom, Bardeen, and ABG black holes. 

However, we found another behavior of the roots of $\mathcal{P}_6(\widetilde{P}_0(\widetilde{r}))$ different to the one shown in figure \ref{SeisRaices}, but with other sets of parameters, for instance, with $\widetilde{a}=0.01,\widetilde{E}=0.8,\widetilde{L}=4$, for small values of $\widetilde{r}$, there are exclusively two roots, which are approaching $\widetilde{E}= -0.8$ very fast as $\widetilde{r}$ increases, specifically $|\widetilde{P}_0(\widetilde{r})-\widetilde{E}|<10^{-4}$ for $\widetilde{r}>6.0$, yet for $\widetilde{r} > 30.0$, $\widetilde{P}_0$ is practically $\widetilde{E}=-0.8$ (and these roots branches exist for all values of $\widetilde{r}$). Nonetheless, other roots appear for $r>\widetilde{r}=14.17$, yet farther away from $\widetilde{E}=-0.8$, these roots converge to $-\widetilde{E}$ as $\tilde{r}$ increases yet, rather slowly; for instance, specifically  at $\widetilde{r} \approx 14.17$, $|\widetilde{P}_0-\widetilde{E}| \approx 0.153$ and at $\widetilde{r} \approx 30.0$, $|\widetilde{P}_0-\widetilde{E}| \approx 0.105$. \\

We shall solely consider those two branches of roots of $\mathcal{P}_6(\widetilde{P}_0)$ that are the closest to $-\widetilde{E}$, one above ($\widetilde{P}^{(a)}_0>-\widetilde{E}$) and one below ($\widetilde{P}^{(b)}_0<-\widetilde{E}$) the value $-\widetilde{E}$; besides, this pair prevails along all the radial coordinate. In all the surveys we have done, generally speaking, we have always found this pair $\widetilde{P}^{(a)}_0$ and $\widetilde{P}^{(b)}_0$ that rapidly approaches $-\widetilde{E}$ and exists in all our numerical radial domain, the other roots may exist in finite sections of the domain, and they do approach $-\widetilde{E}$ yet rather slowly; hence, they will not be considered in our analysis. \\

\begin{figure}[ht]
     \includegraphics[width=\linewidth]{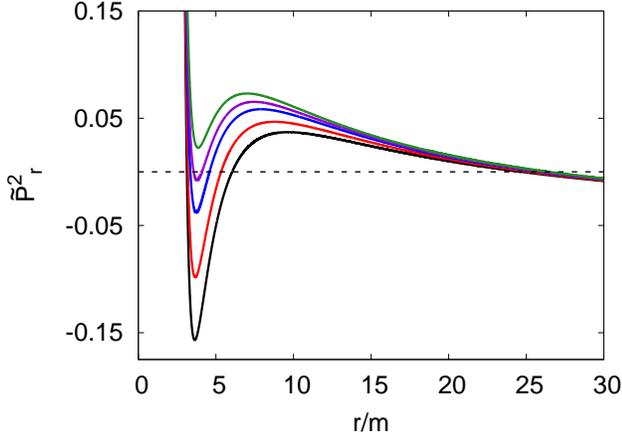}
    \caption{We show the curve $\widetilde{P}_r^2(\widetilde{r})$ for Schwarzschild BH with the parameters $\widetilde{E}=0.97,\widetilde{L}=4$ and $\widetilde{a}=0.2$ (black lowest curve) which has a negative minimum; as $\widetilde{a}$ increases its value $0.3, 0.4, 0.45$ and $\widetilde{a}=0.5$ (green highest curve), the minimum also increases and become positive at $\widetilde{a}_c =0.4585$. For $\widetilde{a}<\widetilde{a}_c$, $\widetilde{P}_r^2(\widetilde{r})$ has three roots and has just one root otherwise.  }
    \label{Pr2}
\end{figure}

Despite the fact that $\widetilde{P}^{(a)}_0$ and $\widetilde{P}^{(b)}_0$ exist throughout the whole range of the radial coordinate, there are regions in the radial coordinate where

\begin{equation}
    \widetilde{P}_r^2 (\widetilde{r}) = \frac{\widetilde{P}_0^2(\widetilde{r})}{f^2(\widetilde{r})} - \frac{1}{f(\widetilde{r})}\left [ \frac{\widetilde{P}^2_{\phi}(\widetilde{r})}{\widetilde{r}^2} + \widetilde{\mathcal{B}}(\widetilde{P}_0(\widetilde{r})) \right ] \nonumber
\end{equation}
\noindent is actually positive and regions where is not, the former case is physically acceptable, the latter is not. Figure \ref{Pr2} shows the curve $\widetilde{P}_r^2(\widetilde{r})$ for Schwarzschild BH for $\widetilde{E}=0.97$, $\widetilde{L}=4.0$ and $\widetilde{a}=0.2$ (black lowest curve), with these parameter's values, one finds that there are three roots of $\widetilde{P}_r^2(\widetilde{r})=0$ that we call $\widetilde{r}_1<\widetilde{r}_2<\widetilde{r}_3$. For $\widetilde{r}<\widetilde{r}_1=3.12732$ and $\widetilde{r}_2=6.04514<\widetilde{r}<\widetilde{r}_3=24.66632$ we found  $\widetilde{P}_r^2(\widetilde{r})$ to be positive. As $\widetilde{a}$ increases its value $0.3, 0.4, 0.45$ and $\widetilde{a}=0.5$ (green highest curve), the roots $\widetilde{r}_1$ and $\widetilde{r}_2$ get closer to each other, collide and then disappear at a certain critical value of $\widetilde{a}_c$. For $\widetilde{a}>\widetilde{a}_c$ there is solely a single root of $\widetilde{P}^2_r(\widetilde{r})=0$. For Schwarzschild, $\widetilde{a}_c \approx 0.4585$. These curves were obtained working with the branch of roots $\widetilde{P}^{(a)}_0(\widetilde{r})$ which is above $-\widetilde{E}$ ($\widetilde{P}^{(a)}_0(\widetilde{r})>-\widetilde{E}$). 

There are similar curves to those presented in figure \ref{Pr2} for the other working examples, Reissner-Nordstrom, Ayon-Beato-Garc\'ia and Bardeen regular BHs. The critical value $\widetilde{a}_c$ where the roots $\widetilde{r}_1$ and $\widetilde{r}_2$ collide is $\widetilde{a}_c=0.689$ for Reissner-Nordstrom, $\widetilde{a}_c=0.9241$ for ABG and $\widetilde{a}_c=0.6436$ for Bardeen BH. Had we employed the lower branch $\widetilde{P}_0^{(b)}(\widetilde{r})< -\widetilde{E}$, one would have found that $\widetilde{P}^2_r(\widetilde{r})$ had just one root $\widetilde{r}_1$ such that, for $\widetilde{r} < \widetilde{r}_1$, $\widetilde{P}^2_r(\widetilde{r})$ is positive and negative otherwise. This fact would mean that close curves (cycles) in the space $(\widetilde{r},\widetilde{P}_r)$ are not allowed. In the end, we avoid working with $\widetilde{P}_0^{(b)}(\widetilde{r})$. Hereafter, we will work with the upper branch $\widetilde{P}_0^{(a)}(\widetilde{r})$ only and consider solely the regions where $\widetilde{P}^2_r(\widetilde{r})$ is positive so that $\widetilde{P}_r(\widetilde{r})$ makes sense. \\
 
\begin{figure}[ht]
     \includegraphics[width=\linewidth]{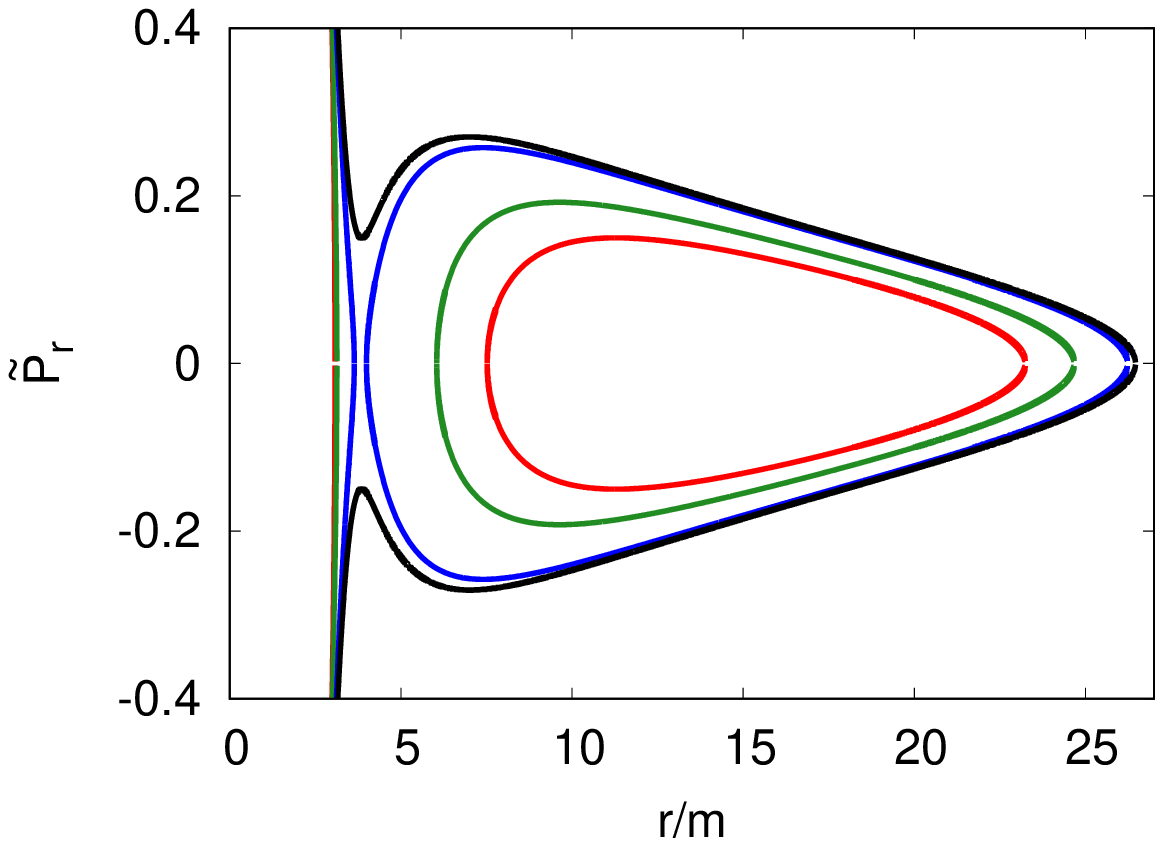}
     \includegraphics[width=\linewidth]{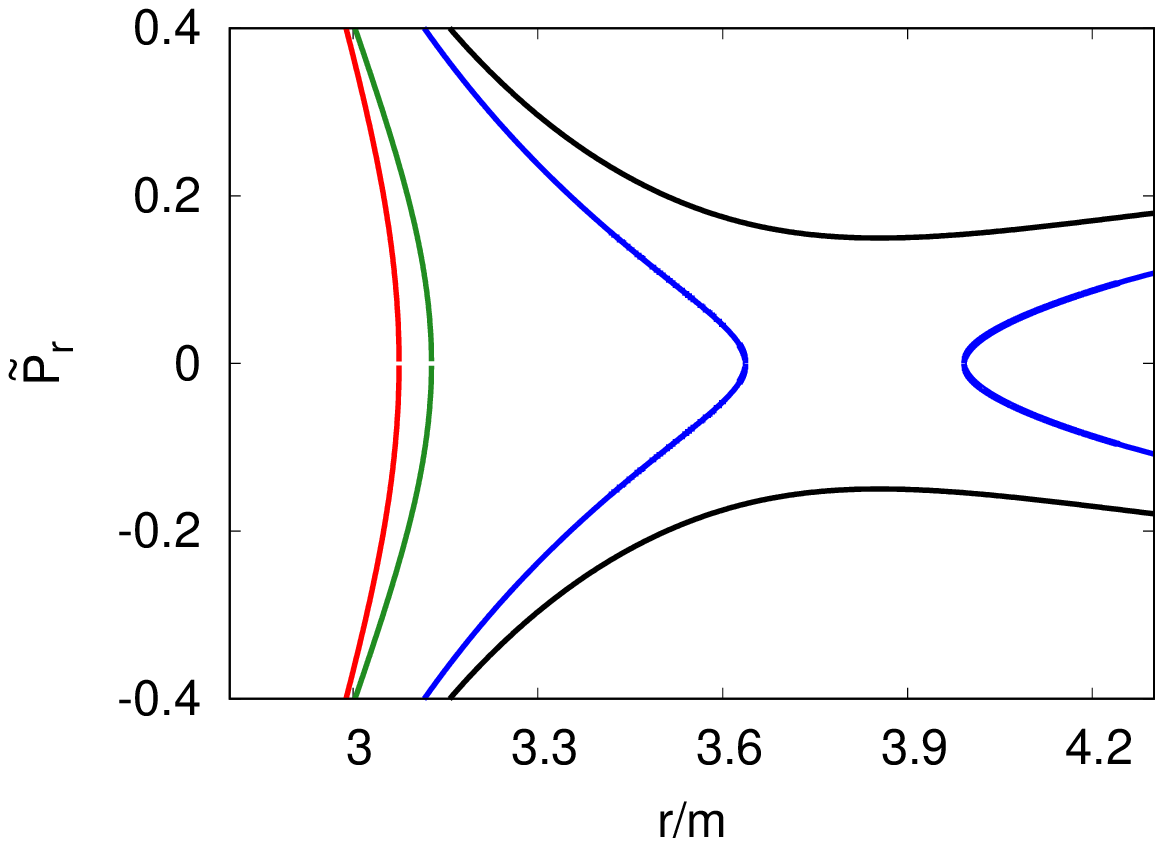}
    \caption{$\widetilde{P}_r(\widetilde{r})$ is shown in the upper panel for $\widetilde{E}=0.97$, $\widetilde{L}=4.0$ and four different values of $\widetilde{a}=0.01, 0.2, 0.45, 0.5$ (red green, blue and black curves) for the Schwarzschild BH. The lower curve is a zoom of the upper plot, showing how the turning points $\widetilde{r}_1$ and $\widetilde{r}_2$ approach each other as $\widetilde{a}$ grows; for $\widetilde{a}=0.5>\widetilde{a}_c=0.458$ (black curve) the have disappeared. }
    \label{Ciclos}
\end{figure}

For the Schwarzschild BH, we plot $\widetilde{P}_r(\widetilde{r})$ with $\widetilde{E}=0.97$, $\widetilde{L}=4.0$ and four different values of $\widetilde{a}=0.01, 0.2, 0.45, 0.5$, red, green, blue and black curves respectively (first plot in figure \ref{Ciclos}). For the first three values of $\widetilde{a}$ the trajectory in phase space $(\widetilde{r},\widetilde{P}_r)$ comprises two sections, one in the interval $(\widetilde{r}_H,\widetilde{r}_1)$ and the other in the interval $(\widetilde{r}_2,\widetilde{r}_3)$ which is a close cycle. As $\widetilde{a}$ climbs, $\widetilde{r}_1$ and $\widetilde{r}_2$ approach each other and $\widetilde{r}_3$ is pushed farther to the right. For $\widetilde{a}=0.5>\widetilde{a}_c=0.458$ (black outer curve) there is only one turning point.   \\

In table \ref{tablaGravi}, we show turning points for $\widetilde{E}=0.97$, $\widetilde{L}=4.0$ and four different values of $\widetilde{a}$, namely $\widetilde{a}=0.01, 0.1, 0.2, 0.3$ for our four spacetime working examples. The behaviour is similar to that found for MPTD spinning particles: as $\widetilde{a}$ increases, the first turning point $\widetilde{r}_1$ increases its value whereas the second turning point $\widetilde{r}_2$ declines and third $\widetilde{r}_3$ climbs up. This feature is followed by the four spacetimes shown in table \ref{tablaGravi}.\\

In order to quantify somehow the effect of the gravimagnetic term on the computation of momenta $\widetilde{P}_{\mu}$, we will heed the difference in the turning points $\widetilde{r}_1$ and $\widetilde{r}_2$ of the cycle trajectories in phase space $(\widetilde{r},\widetilde{P}_r)$. Defining $\triangle_r^{MPTD}= \widetilde{r}_3-\widetilde{r}_2$ and similarly $\triangle_r^{GRAV}$ for the gravimagnetic particles, we may evaluate how the difference $\triangle \equiv \triangle_r^{GRAV}-\triangle_r^{MPTD}$ would vary with $\widetilde{a}$. The results are shown in table \ref{tablaGravi2}. As expected, variations in the radial traveling range of spinning particles increases for gravimagnetic particles compared with those MPTD particles. The difference $\triangle$ is of the order $10^{-5}$ for $\widetilde{a}=0.01$ and of the order $10^{-2}$ for $\widetilde{a}=0.3$

\begin{widetext}

\begin{table}[ht]
    \centering
    \begin{tabular}{|c|c c c|c c c|c c c|c c c|}
    \hline 
    Parameter & \multicolumn{3}{|c|}{Schwarzschild} & \multicolumn{3}{|c|}{Reissner-Nordstrom} & \multicolumn{3}{|c|}{Ayon-Beato-Garcia} & \multicolumn{3}{|c|}{Bardeen} \\
    \hline
    \hline
     $\tilde{a}$ & $\tilde{r}_1$ & $\tilde{r}_2$ & $\tilde{r}_3$ & $\tilde{r}_1$& $\tilde{r}_2$ & $\tilde{r}_3$  & $\tilde{r}_1$ & $\tilde{r}_2$ & $\tilde{r}_3$ & $\tilde{r}_1$ & $\tilde{r}_2$& $\tilde{r}_3$ \\
     \hline
     $0.01$& $3.07475 $ & $7.51981$ & $23.24020$ & $3.05928$ & $7.54306$ & $23.22743$ & $3.04419$ & $7.55225$ & $23.22577$ & $3.05962$ & $7.52908$ & $23.23856$ \\
    \hline
     $0.1$& $3.09107$ & $6.79176$ & $23.95311$ & $3.07411$ & $6.81544$ & $23.94149$ & $3.05790$ & $6.82573$ & $23.94004$ & $3.07478$ & $6.80215$ & $23.95165$ \\
    \hline
     $0.2$& $3.12728$ & $6.04514$ & $24.66632$ & $3.10777$ & $6.07052$ & $24.65569$ & $3.08966$ & $6.08276$ & $24.65440$ & $3.10901$ & $6.05756$ & $24.66503$ \\
  \hline
     $0.3$& $3.19757$ & $5.33184$ & $25.31366$ & $3.17329$ & $5.36136$ & $25.30379$ & $3.15179$ & $5.37716$ & $25.30262$ & $3.17559$ & $5.34806$ & $25.31250$ \\
    \hline
    \hline
    \end{tabular}
    \caption{Values for the turning points $\widetilde{r}_1<\widetilde{r}_2<\widetilde{r}_3$ turning points of gravimagnetic ($\kappa=1$) spinning particles orbiting in our four space-time working examples with $\widetilde{E}=0.97$, $\widetilde{L}=4.0$. As $\widetilde{a}$ increases, $\widetilde{r}_1$ grows and $\widetilde{r}_2$ decreases until the collide and disappear at $\widetilde{a}_c$, whereas $\widetilde{r}_3$ keeps moving to the right. It is for the ABG metric that those changes are the least. The charge value for Reissner-Nordstrom, ABG and Bardeen BH was chosen as $\widetilde{q}=0.1$. }
    \label{tablaGravi}
\end{table}
%
\begin{table}[ht]
    \centering
    \begin{tabular}{|c| c c c| c c c|}
    \hline 
    Parameter & \multicolumn{3}{|c|}{Schwarzschild} & \multicolumn{3}{|c|}{Reissner-Nordstrom} \\
    \hline
    \hline
     $\tilde{a}$ & $\triangle_r^{Gravmag}$ & $\triangle_r^{MPTD}$ & $\triangle$ & $\triangle_r^{Gravmag}$ & $\triangle_r^{MPTD}$ & $\triangle$  \\
     \hline
     $0.01$& $15.72039$ & $15.72036$ & $3(10^{-5})$ & $15.68437$ & $15.68433$ & $4(10^{-5})$  \\
    \hline
     $0.1$& $17.16135$ & $17.15712$ & $4.23(10^{-3})$ & $17.12605$ & $17.12187$ & $4.18(10^{-3})$   \\
    \hline
     $0.2$& $18.62118$ & $18.60114$ & $2(10^{-2})$ & $18.58517$ & $18.56542$ & $1.97(10^{-2})$   \\
    \hline
    $0.3$ &  $19.98182$ & $19.92386$ & $5.79(10^{-2})$ & $19.94243$ & $19.88586$ & $5.65(10^{-2})$ \\
    \hline
    \hline
     & \multicolumn{3}{|c|}{Ayon-Beato-Garcia} & \multicolumn{3}{|c|}{Bardeen} \\
    \hline
    \hline
     $\tilde{a}$ & $\triangle_r^{Gravmag}$ & $\triangle_r^{MPTD}$ & $\triangle$ & $\triangle_r^{Gravmag}$ & $\triangle_r^{MPTD}$ & $\triangle$  \\
     \hline
     $0.01$ &  $15.67352$ & $15.67348$ & $4(10^{-5})$ & $15.70948$ & $15.70945$ & $3(10^{-5})$\\
    \hline
     $0.1$ &  $17.11431$ & $17.11015$ & $4.16(10^{-3})$ & $17.1495$ & $17.14531$ & $4.19(10^{-3})$  \\
    \hline
     $0.2$ & $18.54164$ & $18.55208$ & $1.95(10^{-2})$ & $18.60747$ & $18.58762$ & $1.98(10^{-2})$ \\
    \hline
    $0.3$ &  $19.92546$ & $19.86979$  & $5.56(10^{-2})$ & $19.96444$ & $19.90744$ & $5.7(10^{-2})$\\
    \hline
    \end{tabular}
    \caption{Looking at the cycles in phase space $(\widetilde{r},\widetilde{P}_r)$ as those shown in figure \ref{Ciclos}, with the values of the smallest $\widetilde{r}_2$ and largest $\widetilde{r}_3$ turning points of MPTD ($\kappa=0$) and the gravimagnetic ($\kappa=1$) spinning particles orbiting in our four space-time working examples with $\widetilde{E}=0.97$, $\widetilde{L}=4.0$, one may compute the interval distance a particle may travel $\triangle_r \equiv \widetilde{r}_3-\widetilde{r}_2$. For the values of $\widetilde{a}$ shown, in this table we present how much this distance increase $\triangle \equiv \triangle_r^{Gravmag}-\triangle_r^{MPTD}$ as we consider gravimagnetic particles.  The charge value was chosen as $\widetilde{q}=0.1$. }
    \label{tablaGravi2}
\end{table}

\end{widetext}

\section{Relationship between Velocities and Momenta}

In this final section, we write down explicit expressions for four-velocities $\dot{x}^{\mu}$ (which are not parallel to the momenta) for both cases, the MPTD ($\kappa=0$) and the gravimagnetic
($\kappa=1$) spinning particles moving in spacetimes with a Schwarzschild-like metric we have been working 
\begin{equation}
    ds^2  = -f(r) c^2dt^2 + \frac{1}{f(r)} dr^2 +r^2d\theta^2 + r^2\sin^2\theta d\phi^2 \nonumber
\end{equation}

{\bf Gravimagnetic particles} \\

For this case, the velocity of gravimagnetic spinning particles is given in (\ref{sl4}), which we rewrite here
\begin{eqnarray}
\dot{x}^0 &=& \frac{1}{m_r} \left [ -\frac{P_0}{f(r)} + c Z^0 \right ] \label{x0-dot} \\
\dot{x}^r &=& \frac{1}{m_r} \left [ f(r) P_r + c Z^r \right ] \label{r-dot} \\
\dot{x}^{\phi} &=& \frac{1}{m_r} \left [ \frac{P_{\phi}}{r^2} + c Z^{\phi} \right ] \label{phi-dot}
\end{eqnarray}
where
\begin{equation}
\frac{1}{m_r} = \frac{c}{ \sqrt{ \mu^2 + \frac{1}{16}  \theta\cdot S - c^2Z^2} } \, . 
\end{equation}
In order to obtain the velocities, it is necessary to compute the $Z^{\mu}$ vector components, which in turn employ the Riemann tensor covariant derivatives (the non-vanished covariant derivatives are listed in the appendix) together with the $S^{\mu \nu}$. After a long but straightforward computation, the $Z^{\mu}$ components are attained, they read
\begin{widetext}
\begin{eqnarray}
Z^0 &=& \frac{b}{4c} \left [ f'''(r) P_{\phi}^2 +(r f''(r)-f'(r))\left ( 3 f(r) P_r^2-\frac{P_0^2}{f(r)} \right ) \right ] P_{\phi} \left ( \frac{S^{r \phi}}{P_0} \right )^3 \\
Z^r &=& -\frac{b}{2c} f(r) \left [ r f''(r)-f'(r) \right ] P_{\phi} P_r P_0 \left ( \frac{S^{r \phi}}{P_0} \right )^2 \\
Z^{\phi} &=& -\frac{b}{4c} \left [ f'''(r) P_{\phi}^2+ \left( rf''(r)-f'(r) \right ) \left [ f(r) P_r^2-\frac{P_0^2}{f(r)} \right ] \right ] P_0 \left ( \frac{S^{r\phi}}{P_0} \right )^3
\end{eqnarray}
where the scalar  $b=(8\mu^2 + \theta\cdot S)^{-1}$.  Combining (\ref{Theta-S})  with (\ref{S-r0}) and (\ref{S-phi0}) we have
\begin{equation}
    \theta \cdot S = 2 \left [ f''(r) P_{\phi}^2+r f(r) f'(r) P_r^2- r\frac{f'(r)}{f(r)} P_o^2 \right ] \left ( \frac{S^{r\phi}}{P_0}\right )^2
\end{equation}
$Z^{\theta}$ is omitted since we are considering motion in the equatorial plane solely. $Z^2$ is also needed
\begin{equation}
    Z^2 \equiv Z_{\mu}Z^{\mu}= -f(r) \left ( Z^0 \right )^2+\frac{\left ( Z^r \right )^2}{f(r)}+r^2 \left ( Z^{\phi} \right )^2
\end{equation}
\end{widetext}

 {\bf MPTD particles} \\
 
The velocity components in this case are given in (\ref{sl0-7}), which we rewrite here
\begin{eqnarray}
    \dot{x}^0&=&\frac{c \left[ -\frac{P_0}{f(r)} + \bar{a} \xi^0 \right ]}{\sqrt{\mu^2-\bar{a}\xi^2}} \\
    \dot{x}^r &=& \frac{c \left[ f(r) P_r + \bar{a} \xi^r \right ]}{\sqrt{\mu^2-\bar{a}\xi^2}} \\
     \dot{x}^{\phi} &=& \frac{c \left[ \frac{P_{\phi}}{r^2} + \bar{a} \xi^{\phi} \right ]}{\sqrt{\mu^2-\bar{a}\xi^2}} 
\end{eqnarray}
where $\bar a= 2(16\mu^2 + \theta\cdot S)^{-1}$. $\xi^{\mu}$ is defined as  
\begin{equation}
    \xi^{\mu} \equiv S^{\mu \alpha} \theta_{\alpha \nu} P^{\nu}
\end{equation}
which turn out to be
\begin{eqnarray}
\xi^0 &=& -\frac{1}{f(r)} \left ( \frac{f'(r)}{r}-f''(r) \right ) P_0 P_{\phi}^2 \left ( \frac{S^{r\phi}}{P_0}\right )^2 \label{x0-0-dot}\\
\xi^r &=& f(r) \left ( \frac{f'(r)}{r}-f''(r) \right ) P_r P_{\phi}^2 \left ( \frac{S^{r\phi}}{P_0}\right )^2 \label{r-0-dot}\\
\xi^{\phi} &=& -\frac{1}{f(r)} \left ( \frac{f'(r)}{r}-f''(r) \right ) \left [ f^2(r) P_r^2-P_0^2 \right ] P_{\phi} \left ( \frac{S^{r\phi}}{P_0}\right )^2 \label{phi-0-dot} \nonumber \\
\end{eqnarray}
and 
\begin{equation}
    \xi^2 = -f(r) \left ( \xi^0 \right )^2 + \frac{\left ( \xi^r \right )^2}{f(r)}+r^2 \left ( \xi^{\phi} \right )^2
\end{equation}

\section{Conclusions and final remarks}

In this work, we have studied the equations of motion of spinning particles with gravimagnetic interaction (\ref{g-m}).  We focused on two choices of gravimagnetic moment $\kappa$; one corresponds to $\kappa=0$, where the equations are fully comparable with the MPTD equations (see section \ref{comparation MPTD}). The other choice is  $\kappa=1$, in this case, the equations have a correct behavior in ultra-relativistic limit (see \cite{Deriglazov:2017jub, DWGR2016, Deriglazov:2015zta, Deriglazov:2014zzm}). The study has been done in a geometrical background with a metric of the form 
$ds^2=-f(r) c^2 dt^2+ \frac{dr^2}{f(r)} +r^2(d\theta^2+\sin^2{\theta} d\phi^2)$,
specifically with Schwarzschild, Reisser-Nordstrom black holes as well as Ayon-Beato-Garcia and Bardeen regular spacetimes. Yet any spacetime with this form of the metric, can be included in our study. The equations for $\kappa=0$ (MPTD particles) and $\kappa=1$ (Gravimagnetic particles) were fully analysed on the equatorial plane and we proved that the set of differential equations for the momenta and the non-vanishing components of $S^{\mu \nu}$ might be replaced by the set of algebraic equations (\ref{f1})-(\ref{f4}). For $\kappa=0$, explicit analytical expressions for the momenta $P_\mu(r)$ and the relevant components $S^{\mu \nu}$ were presented. 

For $\kappa=1$, the equation for $P_0$ turned out to be a single polynomial of sixth degree $\mathcal{P}_6(P_0)$, which in principle may have six real roots. In order for a root $P_0^{*}$ to be physically acceptable, it must lie in the interval $(P_0^{-},P_0^{+})$ where $P_0^{\pm}$ are roots of the quadratic equation for $P_0$ given by (\ref{BP0}); for our four spacetime examples, we numerically verified that there are always roots in that interval, at least two. A numerical algorithm was presented in section III, to construct solutions for the momenta $P_{\mu}(r)$ and $S^{\mu \nu}(r)$ for a given set of the particles parameters $E$, $L$ and $a$ (energy, angular momentum and spin). 

It was numerically shown that, for close trajectories (cycles) in phase space $(r,P_r)$, the radial traveling range of spinning particles increases for gravimagnetic particles compared with those MPTD particles. 

The expressions found in the previous section for velocities, might be employed to study the orbits for MTDP and gravimagnetic particles using any of the spacetimes studied in this article. It might be interesting to carry out similar studies of spacetimes with a metric of the form

    $ds^2=-A(r) c^2 dt^2+ \frac{dr^2}{B(r)} +r^2(d\theta^2+\sin^2{\theta} d\phi^2)$

\noindent which includes the case for boson stars and boson-fermion stars \cite{Colpi, Liebling:2012fv, Valdez-Alvarado:2020vqa, Valdez-Alvarado:2012rct, Degollado}. This is a work in progress that will be reported somewhere else.
\begin{acknowledgments}
\noindent WGR: This study was financed in part by the Funda\c c\~ao Carlos Chagas Filho de Amparo \`a Pesquisa do Estado do Rio de Janeiro (FAPERJ) -Finance Code 211.622/2019. RB acknowledges partial financial support from CIC-UMSNH and SNI-Conacyt. The work of AAD has been supported by the Brazilian foundation CNPq (Conselho Nacional de Desenvolvimento Cient\'ifico e Tecnol\'ogico - Brasil),  and by Tomsk State University Competitiveness Improvement Program.
\end{acknowledgments}

\section{Appendix}
Non vanishing components of the Riemann tensor, $R_{\mu\nu\alpha\beta}$:
\begin{eqnarray} 
	R_{ 0 r0r} &&= \frac{1}{2}\frac{d^2f(r)}{dr^2} \, ,\nonumber \\
         R_{ 0 \theta 0 \theta} &&= \frac{rf(r)}{2} \frac{df(r)}{dr} \, ,\nonumber \\ 
         R_{ 0\phi 0 \phi} &&=  \frac{rf(r)}{2} \frac{df(r)}{dr}\sin^2(\theta)  \, ,\nonumber \\
\end{eqnarray}
\begin{eqnarray}
         R_{ r \theta r \theta} &&= -\frac{r}{2f(r)} \frac{df(r)}{dr} \, , \nonumber\\
         R_{ r \phi r \phi} &&=  -\frac{r}{2f(r)} \frac{df(r)}{dr}\sin^2(\theta)\, ,\nonumber \\
         R_{ \theta \phi\theta\phi} &&= -r^2\sin^2(\theta) - r^2\sin^2(\theta)f(r) \, .\nonumber
\end{eqnarray}

Non vanishing covariant derivatives components of the Riemann tensor:\vspace{0.5cm}

\noindent $\nabla_r R_{\mu\nu\alpha\beta}$:
\begin{eqnarray}
\nabla_r R_{0r0r} &=&\frac{1}{2} \frac{d^3 f(r)}{dr^3}\, ,\nonumber \\
\nabla_r R_{0\theta 0\theta} &=& \frac{1}{2} \left( -f(r) \frac{d f(r)}{dr}+rf(r) \frac{d^2f(r)}{dr^2}  \right)\, , \nonumber\\
\nabla_r R_{0\phi 0\phi} &=& -\frac{1}{2} f(r) \frac{d f(r)}{dr}\sin^2{\theta} \nonumber \\
&& \nonumber\\
&& + \frac{1}{2} r f(r) \frac{d^2f(r)}{dr^2} \sin^2{\theta}\, ,\nonumber  \\
\nabla_r R_{r\theta r\theta} &=&-\frac{1}{2f(r)} \left( r\frac{d^2f(r)}{d^2r} - \frac{df(r)}{dr} \right)\, , \nonumber \\
\nabla_r R_{r \phi r \phi} &=& -\frac{\sin^2(\theta)}{2f(r)} \left( r\frac{d^2f(r)}{d^2r} - \frac{df(r)}{dr} \right) \, ,\nonumber \\
\nabla_r R_{\theta \phi \theta \phi} &=& -r\sin^2(\theta)\left(r\frac{df(r)}{dr}-2f(r) +2 \right)\, .\nonumber
\end{eqnarray}

\noindent $\nabla_\theta R_{\mu\nu\alpha\beta}$: 
\begin{eqnarray}
\nabla_\theta R_{0r0\theta} &=& - \frac{f(r)}{2} \left(\frac{df(r)}{dr} - r\frac{d^2f(r)}{d^2r}  \right)\, , \nonumber \\
\nabla_\theta R_{r\phi \theta\phi} &=& \frac{-\sin^2(\theta)}{2}\left(r\frac{df(r)}{dr} -2f(r) + 2 \right)\, .\nonumber
\end{eqnarray}

\noindent $\nabla_\phi R_{\mu\nu\alpha\beta}$: 
\begin{eqnarray}
\nabla_\phi R_{0r0\phi} &=& - \frac{f(r)\sin^2(\theta)}{2} \left(\frac{df(r)}{dr} -r\frac{d^2f}{dr^2} \right)\, ,\nonumber \\
\nabla_\phi R_{r\phi \theta\phi} &=& -\frac{r\sin^2(\theta)}{2}\left(r\frac{df(r)}{dr}-2f(r)+2\right). \nonumber\\ \nonumber
\end{eqnarray}

$\theta_{\mu\nu}=R_{\mu\nu\alpha\beta}S^{\alpha\beta}$ - components.

$\theta_{\mu\nu}= 2R_{\mu\nu \ i0} S^{i0} + R_{\mu\nu \ ik} S^{ik}$

$\theta_{\mu\nu}\dot x^\nu$:
\begin{eqnarray}
\theta_{0\nu} \dot x^\nu=2\left[ R_{r0r0} S^{r0} \dot r + R_{\theta 0\theta 0} S^{\theta 0} \dot \theta + R_{\phi 0\phi 0} S^{\phi 0} \dot \phi \right]\, , \nonumber \\
\theta_{r\nu} \dot x^\nu=2\left[ R_{r0r0} S^{r0} \dot x^0 + R_{r \theta r \theta } S^{r \theta} \dot \theta + R_{r \phi  r \phi } S^{r \phi } \dot\phi \right]\, , \nonumber\\
\theta_{\theta\nu} \dot x^\nu=2\left[ R_{\theta0 \theta0} S^{\theta 0} \dot x^0 - R_{r \theta r\theta } S^{r \theta} \dot r + R_{\theta\phi \theta\phi } S^{\theta\phi } \dot \phi \right]\, , \nonumber\\
\theta_{\phi\nu} \dot x^\nu=2\left[ R_{\phi 0\phi 0} S^{\phi 0} \dot x^0 - R_{r\phi r\phi} S^{r \phi} \dot r - R_{\theta\phi \theta\phi } S^{\theta\phi} \dot \theta \right]\, .\nonumber
\end{eqnarray}

%


\end{document}